\documentclass[9pt]{article}

\usepackage[
backend=biber,
style=numeric,
sorting=none,
maxbibnames=99, minbibnames=99
]{biblatex}
\usepackage{graphicx}
\usepackage{amsmath}
\usepackage{amsthm}
\usepackage{amssymb}
\usepackage{cancel}
\usepackage{comment}
\usepackage{subfig}
\usepackage{lipsum}
\usepackage{color}
\usepackage{multirow}
\usepackage{authblk}
\usepackage[a4paper, total={6in, 8in}]{geometry}

\usepackage{hyperref}

\newcommand{\eps}{\varepsilon}

\title{Beating of eukaryotic flagella via Hopf bifurcation of a system of stalled molecular motors}

\author[1]{Irene Anello}
\author[2]{François Alouges}
\author[1,3]{Antonio De Simone}
\affil[1]{MathLab, Scuola Internazionale Superiore di Studi Avanzati (SISSA), Via Bonomea 265, Trieste (TS), 34136, Italy}

\affil[2]{Centre Borelli, ENS Paris-Saclay, CNRS, Université Paris-Saclay and Institut Universitaire de France, 4, avenue des Sciences, Gif-sur-Yvette, 91190, France}

 \affil[3]{The BioRobotics Institute, Scuola Superiore Sant'Anna, Viale R. Piaggio 38, Pontedera (PI), 56025, Italy}
\date{}
\begin{document}

\maketitle
\tableofcontents
\newpage

\begin{abstract}
The modeling of the beating of cilia and flagella in fluids is a particularly active field of study, given the biological relevance of these organelles. Various mathematical models have been proposed to represent the nonlinear dynamics of flagella, whose motion is powered by the work of molecular motors attached to filaments composing the axoneme. Here, we formulate and solve a nonlinear model of activation based on the sliding feedback mechanism, capturing the chemical and configurational changes of molecular motors driving axonemal motion. This multiscale model bridges microscopic motor dynamics with macroscopic flagellar motion, providing insight into the emergence of oscillatory beating. We validate the framework through linear stability analysis and fully nonlinear numerical simulations, showing the onset of spontaneous oscillations. To make the analysis more comprehensive, we compare our approach with two established sliding feedback models.
\end{abstract}

\section{Introduction}\label{Introduction}
Eukaryotic flagella exhibit characteristic beating patterns that enable propulsion in viscosity-dominated environments \cite{Purcell, lauga2009hydrodynamics}. This motion is powered by molecular motors that operate within the axoneme, a cytoskeletal structure conserved across species. Due to the importance of flagellar beating in various biological functions, understanding the mechanisms underlying motor coordination in the axoneme remains a central question in biophysics \cite{gaffney2011mammalian, smith2004radial, heuser2009dynein, Velho2021, EukFlagella}. 

The axoneme has a 9+2 structure, with nine microtubule doublets surrounding a central pair of microtubule doublets, as shown in Figure~\ref{fig:axoneme}. Each doublet consists of an
A-tubule and a B-tubule. Between these doublets are molecular motor
proteins (dyneins), which anchor their tails to the A-tubule while their heads are free to
bind and release the B-tubule of the subsequent doublet \cite{lin2018asymmetric}. Since the microtubules are polar filaments, this ATP-powered binding action creates movement towards the minus end of the microtubule,
resulting in the relative sliding of adjacent doublets. The axoneme is held together by passive elastic elements, such as radial spokes proteins and nexin cross-linkers, that, together with constraints at the base of the flagellum, transform the sliding of microtubules into bending of the axoneme.

A major focus of flagellar research is to understand how the forces generated by dynein acting at the microscopic scale coordinate to generate bending waves in the flagellum. Various mathematical models have been proposed to describe these interactions. A seminal study in this field was conducted by Machin \cite{machin1958wave}, who demonstrated the importance of internal activation within a flagellum as a mechanism to sustain oscillations. Interestingly, the equation first derived there has recently been rederived in \cite{howard2022predicting} using a molecular mechanics approach.

Different hypotheses on the interaction between motors and microtubule pairs lead to different feedback (or control) mechanisms, including sliding feedback \cite{brokaw1985computer,brokaw1971bend,gallagher2023axonemal,geyer2022ciliary,mondal2020internal}, curvature feedback \cite{camalet2000generic,riedel2007molecular,oriola2017nonlinear}, and geometric clutch \cite{lindemanngeometric1994, lindemann1994model, bayly2014equations}. Within these models, feedback dynamics can be categorized as linear or non-linear. Recently, fully non linear models have been shown to effectively describe the behavior of small flagella. In particular, while linear curvature feedback was initially considered ideal for fitting \textit{Chlamydomonas} data \cite{sartori2016dynamic}, the non-linear sliding feedback model proposed by \cite{cass2023reaction}, which incorporates the attachment and detachment of antagonistic molecular motors, provides a better fit.
A review of the three type of control models can be found in \cite{sartori2016dynamic, bayly2015analysis}. It is worth noting the existence of flagellar beating models that do not depend on mechanical feedback between motors and filaments but instead reproduce oscillatory flutter instabilities through an external follower force at the free end of an active filament \cite{Keaveny2024bif,ling2018instability}.

This study introduces a fully non-linear model for flagellar waveforms which belongs to the family of sliding feedback models. This model couples a motor model obtained generalizing \cite{ camalet2000generic,julicher1995cooperative} with the equation of motion for the flagellum modeled  as a pair of filaments in a plane. We refer to this framework as the $\mu$-\textit{chemoEH} model\footnote{micro-chemo-Elasto-Hydrodynamic.}. Respecting the symmetric structure of the axoneme, this model aims to elucidate the tug-of-war scenario of antagonistic motors between filaments at the microscopic scale \cite{alouges2024mathematicalmodelsflagellaractivation}. Additionally, it preserves the non-linearity given by the transport term originally proposed in \cite{julicher1995cooperative,julicher1997modeling}, which uses a non-linear reaction-advection equation to model the distribution of molecular motors in space.

In this scenario, the beating of the eukaryotic flagellum is explained as an oscillatory (Hopf) bifurcation.
When there is enough ATP, the stalled equilibrium configuration of molecular motors becomes unstable, and results in an oscillatory tug-of-war causing the alternating relative sliding of filaments.
This work contains two main original contributions: firstly, the exploration of the two-state model in the non-linear regime; and secondly, the investigation of the distribution of probabilities (that motors are bound/unbound) at the microscopic scale, when large bending deformations arise at the macroscopic scale.

In Section~\ref{sec:nonlinear_continuum} we derive the equation of motion for the flagellum, which is modeled as two filaments that bend in a plane. Subsequently, we present three nonlinear chemical models: the $\mu$-chemoEH, the cubic model (a simplified version of the former), and the chemoEH model \cite{oriola2017nonlinear}. Then, we conduct the linear analysis of the $\mu$-chemoEH on long flagella and  corroborate it with nonlinear simulations in the vicinity of the bifurcation point. Finally, we present and discuss a numerical comparison between the three models in the large-amplitude regime in the context of short flagella.

\section{Non linear continuum equations}\label{sec:nonlinear_continuum}
\subsection{Kinematics}
\begin{figure}
    \centering
    \includegraphics[width=0.5\linewidth]{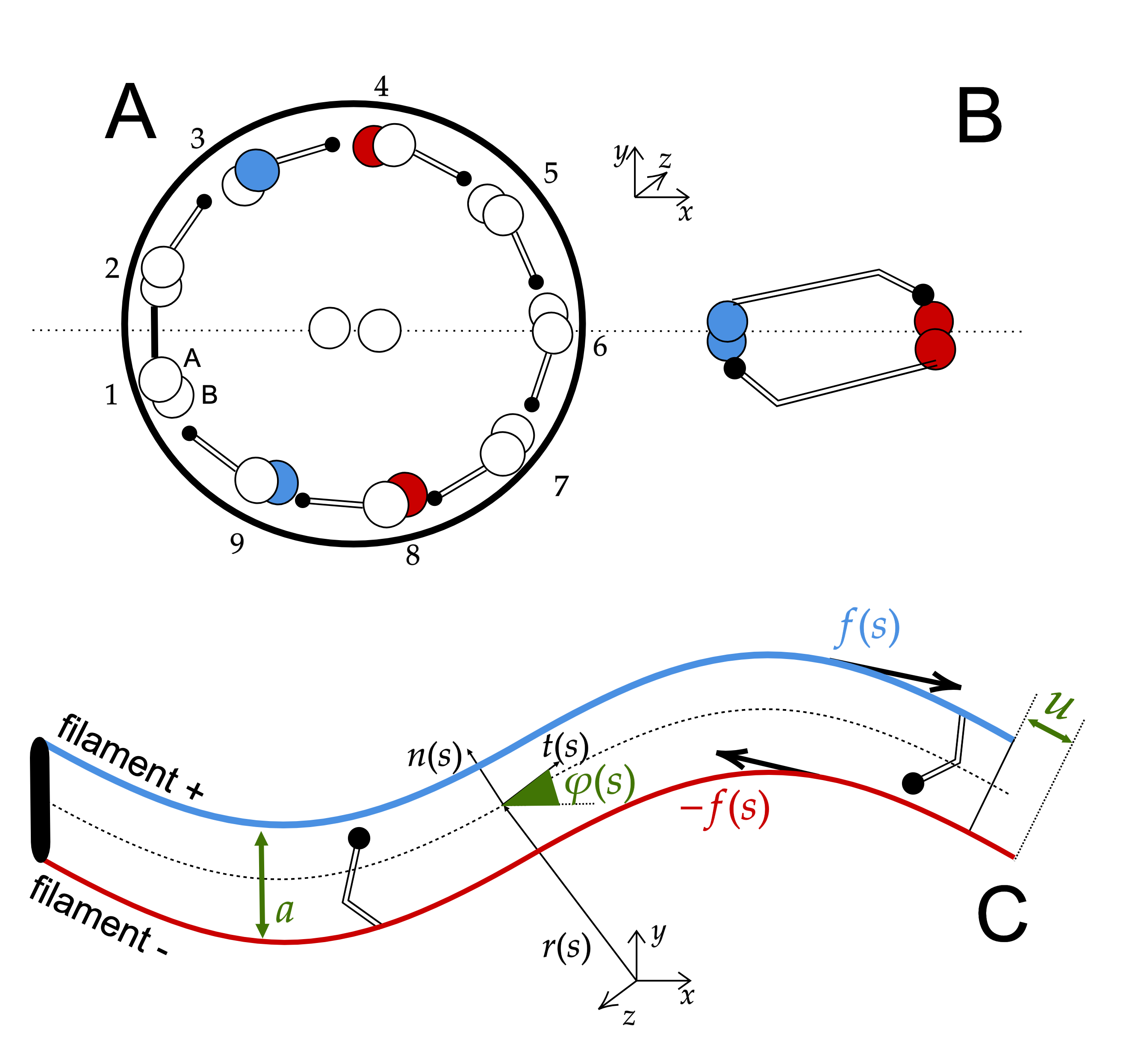}
    \caption{From the axoneme to the two rows of molecular motors. ({A}) Cross section of the axoneme when viewed from the base to tip, with numbering taken from \cite{sartori2016dynamic}. ({B}) projection of the axoneme onto its bending plane. (C) Flagellum composed by two filaments fixed at the base.}
    \label{fig:axoneme}
\end{figure}

We use a simplified model for the axoneme, extensively used in literature \cite{camalet2000generic, riedel2007molecular, sartori2016dynamic, cass2023reaction}, where its structure reduces to two microtubule doublets, as in Figure~\ref{fig:axoneme}. 

The microtubules are modeled as a pair of elastic and inextensible filaments with same length $L$ and at fixed distance $a \ll L$. The centerline curve is parameterized as $s \to \mathbf{r}(s) \in \mathbb{R}^2$, where $s \in [0,L]$ is the arc-length. The two filaments are then parametrized by 
\begin{equation}\label{eq: r+,r-}
    \mathbf{r}_+(s)= \mathbf{r}(s)+a \mathbf{n}(s)/2, \quad \mathbf{r}_-(s)= \mathbf{r}(s)-a \mathbf{n}(s)/2.
\end{equation}

The local tangent vector $\mathbf{t}(s)$ is 
\begin{equation}\label{eq: r'=t}
     \mathbf{t}(s) = \mathbf{r}'(s).
\end{equation}
From here on the symbol $'$ indicates the arc-length derivative $\partial/\partial_s$. 
We denote by $\varphi(s)$ the angle between $\mathbf{t}(s)$ and the x-axis, such that $\mathbf{t}=(\cos \varphi,\sin \varphi)$, and by $\kappa(s) = \varphi'(s)$ the local curvature, such that $\mathbf{n}' = -\kappa\mathbf{t}$.

The bending of the pair of filaments is related to their relative sliding. The sliding displacement at position $s$ along the neutral line is given by the difference between the total arc-lengths along the two filaments up to the points $\mathbf{r}_+(s)$ and $\mathbf{r}_-(s)$, namely 
\begin{equation}\label{eq: sliding def}
    u(s) = u(0)+ \int_0^s (|\mathbf{r}_+'(\sigma)|- |\mathbf{r}_-'(\sigma)|)\,d\sigma = u(0) + a(\varphi(s)-\varphi(0)),
\end{equation}
where $u(0)$ is the basal sliding of one filament with respect to the other, and we consider it null for simplicity. We remark that in deducing the last identity \eqref{eq: sliding def} from the tangent and normal differential relations, we use the assumption $ 1 - a \kappa > 0$. Notice that by differentiating \eqref{eq: sliding def} we obtain
\begin{equation}\label{eq:u'=ak}
    u' = a \varphi'= a \kappa.
\end{equation}

\subsection{Force and torque balance equations}
The center filament is subject to internal forces $\mathbf{R}(s)$ and torque $M(s)$ that act at position $s$. The internal forces are balanced by the external hydrodynamic drag per unit length $\mathbf{f}_{fl}(s)$ through the local force balance equation
\begin{equation*}
    \mathbf{R}'(s) + \mathbf{f}_{fl}(s) =0.
\end{equation*}
Since we are dealing with a slender body at low Reynolds number, we use the resistive force theory (RFT,\cite{gray1955propulsion}) limit and express the external drag as a function of the centerline velocity $\mathbf{f}_{fl} = - \xi_n  \dot r_n \mathbf{n} - \xi_t \dot r_t \mathbf{t}$, where $\xi_n$ and $\xi_t$ are the normal and tangential friction coefficients, while $\dot{r}_n$ and $\dot{r}_t$ are the normal and tangential components of the centerline velocity. The dot indicates the time derivative $\partial/\partial_t$.

The local torque balance equation reads
\begin{equation}\label{eq: local moment balance}
    M'(s)+T(s)+m(s)=0,
\end{equation}
where $M(s)= B \varphi'(s)$ is the bending moment proportional to the curvature with bending stiffness $B$, $T(s)$ is the shear stress, defined as \[T(s) = \mathbf{R}(s) \cdot \mathbf{n}(s),\]
and $m(s)$ is the distributed moment density. 

 We impose clamped-free boundary conditions
\begin{equation} \label{BC}
\varphi(0)=0,\quad \dot{\mathbf{r}}(0)=0,\quad \mathbf{R}(L)=0, \quad M(L)=0.
\end{equation}
Since the position depends on the tangent angle by
\begin{equation*}
    \mathbf{r}(s,t) = \mathbf{r}(0,t) + \int_0^L\begin{pmatrix}
         \cos(\varphi(\sigma,t)) \\
         \sin(\varphi(\sigma,t))
    \end{pmatrix}
    \, d\sigma,
\end{equation*}
and given that no force is applied at the tip of the flagellum, the shear force depends on the tangent angle by writing $T(s) = \left( \int_s^L \mathbf{f}_{fl}(\sigma) \, d\sigma \right) \cdot \mathbf{n}(s)$. Finally, assuming $m(s)$ is known, we can solve equation \eqref{eq: local moment balance} for the tangent angle $\varphi$.

\subsection{Modeling activity}
In the following, the distributed moment density $m$ will be written as the product $m=af$, where $f=f(s,t)$ is a force density acting at position $s$ in opposite directions on the two filaments. The force $f$ is the sum of a passive part $f_p$ and an active part $f_a$, $f=f_a+f_p$. The passive part is defined as the sum of an elastic and a viscous resistance to sliding, while the active part is an unknown of the problem. To close the system we need then a further equation for the active part, which will encode a feedback between the activity of the motors and the sliding $u$, or its velocity $\dot u$. In the following, we give the equations in this form
\begin{equation} \label{eq: microscopic system}
    \left\{\begin{array}{ll}
    \dot {X} = F( X, \dot u),\\
             f_a = G(X),  \\
         f_p = H(u,\dot u) \\
   \end{array} \right.
\end{equation}
where $F$ is a vector, $G$ and $H$ are  scalars and $X$ is a vector of variables related to the microscopic activity inside the flagellum.

The active force $f_a$ is exerted by the motors and powers the sliding $u$ of the filaments which in turn determines their bending via equation \eqref{eq:u'=ak}.
In the following sections, we describe three alternative versions of system \eqref{eq: microscopic system}.

\section{Chemo-mechanical models}

\subsection{The $\mu$-chemoEH model}
The activity inside the flagellum is modeled by considering molecular motors anchored to the upper filament ($\mathbf{r}_+$) as well as those anchored to the lower filament ($\mathbf{r}_-$), as shown in Figure~\ref{fig:axoneme}. Due to the polarity of the filaments, the molecular motors attach to and detach from the opposite filament relative to the one they are anchored to, with a directional motion to the minus end, creating sliding between the filaments. In this section we present the $\mu$-chemoEH model, a generalization of the microscopic model illustrated in \cite{camalet2000generic}, that relates the force generated by the motor-filament interaction to the relative sliding of the filaments. A description of this model is provided in \cite{alouges2024mathematicalmodelsflagellaractivation}, where the microscopic system was not coupled with the equations of motion for the filaments. In that work, the model was referred to as the two-row model for reasons that will become evident later.

Given the need to relate the macroscopic dynamics of microtubules and swimming flagella with the microscopic mechanism of motors, we introduce different scales. The largest one is defined by the arc-length $s \in [0,L]$, while the smallest is the uniform spacing $\delta$ between molecular motors. Within each small rod segment $(s, s+ ds)$, we consider the presence of microscopic rigid segments of length $ \ell$, with $ \delta \ll \ell \ll L $, reflecting the periodic microtubule structure and containing a large number $N$ of motors. These segment, defined as tug-of-war units \cite{oriola2017nonlinear}, are described by the spatial coordinate $ \xi \in [0, \ell] $. We focus on a specific point $s \in [0,L]$ and, for simplicity, we omit the dependence on $s$ of all quantities in the rest of this subsection. 

At the $\ell$-scale, a stochastic two-state model is employed to model the activity of molecular motors. Each motor exists in one of two states: bound (state 1) to the filament or unbound (state 2) from it. Transitions between these states are driven by the consumption of chemical energy in the form of ATP. This energy is converted into mechanical work by going from the unbound state to the bound one. Each state has an associated potential, $W_1(\xi)$ and $W_2(\xi)$, whose difference is defined as $\Delta W = W_1-W_2$, while the switch between states is described by transition rates $\omega_{12}(\xi)= \omega_1(\xi)$ and $\omega_{21}(\xi)= \omega_2(\xi)$, representing the probability per unit time for a motor to go from state $1$ to state $2$ and vice versa. Since ATP drives these transitions, the rates depend on its concentration, denoted as $\Omega$. 

Referring to Figure~\ref{fig:axoneme},
the relative displacement of the upper filament with respect to the lower filament, is defined by $u(t)$, with velocity $\dot u(t)$. Let us define $P^+(\xi,t)$ as the probability density that a motor anchored to upper filament is in state $1$ at position $\xi$ and at time $t$. The motor force density per macroscopic unit length exerted by the motors anchored to the lower filament on the upper filament is defined as $-f^+$ and reads
\begin{equation}\label{eq:fplus}
   f^+(t) = -\rho N \int_0^{\ell} P^+(\xi,t) \partial_{\xi} \Delta W (\xi)d\xi,
\end{equation}
where $\rho$ is the density of tug-of-war segments in the filament. Note that $\rho N$ is the number density of molecular motors along the flagellum, denoted in \cite{camalet2000generic} as $\rho$. We follow the force-sign convention of \cite{howard2022predicting}. Similarly, $Q(\xi,t)$ is defined as the probability density that a motor anchored to the lower filament is in state $1$ but located at $\xi$ at time $t$. For simplicity, we introduce the notation $P^-(\xi+u(t),t)=Q(\xi,t)$, and we define as
\begin{equation}\label{fmin}
   f^-(t) = -\rho N \int_0^{\ell} P^-(\xi,t) \partial_{\xi} \Delta W (\xi)d\xi
\end{equation}
the correspondent motor force for the lower filament.

The total active force density $f_a$ felt by the upper filament is therefore
\begin{equation}\label{fatworow}
  f_a= f^+ + (-f^-).   
\end{equation}
Viscous and elastic elements, represented by $\lambda$ and $K$ respectively, oppose the motion induced by the motors. The force balance on the upper filament can be then written as
\begin{equation}\label{eq: force balance equation + filament}
    0=f_{\text{ext}}(t) -2 K u(t) -2 \lambda \dot u(t) + f_a(t),
\end{equation}
where $f_{ext}(t)= - f(t)$.

For $\xi \in [0, \ell]$ and $t>0$, the complete PDE system describing the model is
\begin{align} \label{eq:PDE motors}
    \left\{\begin{array}{lll}
    &\displaystyle   \frac{\partial P^+}{\partial t}(\xi,t) + \dot u(t) \frac{\partial P^+}{\partial \xi}(\xi, t)  = -( \omega_1(\xi) +\omega_2(\xi)) P^+(\xi, t) + \frac{\omega_2(\xi)}{\ell},\\[10pt]
   &\displaystyle \frac{\partial P^-}{\partial t}(\xi,t) - \dot u(t) \frac{\partial P^-}{\partial \xi}(\xi,t)  = -( \omega_1(\xi) +\omega_2(\xi)) P^-(\xi,t) + \frac{\omega_2(\xi)}{\ell}, \\
   & f(t)= - \lambda \dot u(t) - K u(t) - {\rho N}\int\limits_0^{\ell} (P^+(\xi,t)- P^-(\xi,t))\partial_{\xi} \Delta W (\xi)d\xi.\\
    \end{array}
    \right.
\end{align}

\begin{figure}[ht!]
    \centering
    \includegraphics[width=0.6\linewidth]{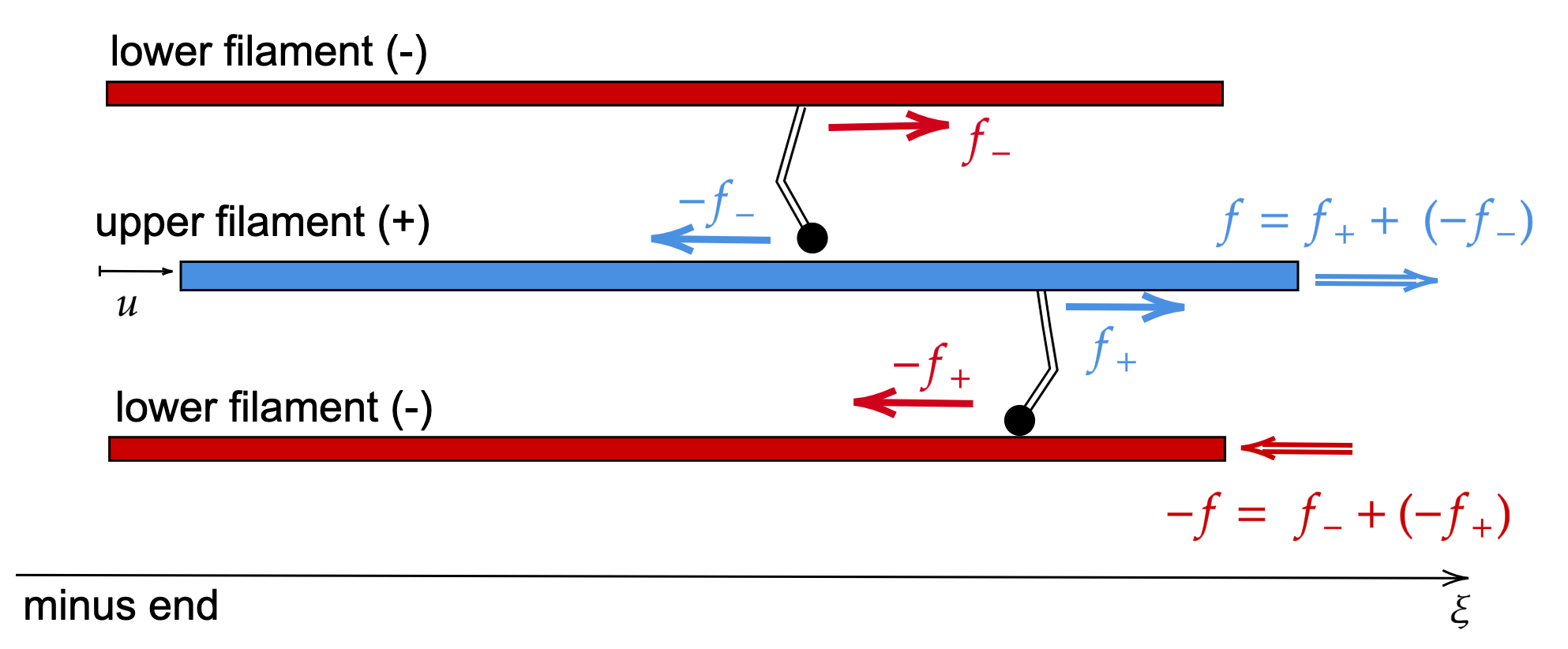}
        \caption{Unfolding of the flagellum two-filament structure in~\ref{fig:axoneme}(B) and (C). The red filaments are the same filament, we refer to it as the lower filament (-). The two rows of competing molecular motors exert forces $f$ on the upper filament and $-f$ on the lower filament.}
    \label{fig: unfolding}
\end{figure}

The unfolding of the two filaments describing the flagellum (Figure~\ref{fig:axoneme}B and ~\ref{fig:axoneme}C), creates two rows of molecular motors, as can be seen in Figure~\ref{fig: unfolding}. 
We chose this framework to describe the axoneme, rather than the simpler one-row model where the motors are attached only to the upper filament as in \cite{camalet2000generic}, to capture the intuitive mechanics at play.
When $\partial_{\xi}\Delta W<0$ we have exactly the situation presented in Figure~\ref{fig:axoneme}(C), and the two rows of antagonistic motors compete through the forces, opposite in sign, $f^+$ and $f^-$. This tug-of-war framework implies that, when $\dot u=0$ and the system is in equilibrium with probability density $P^+_0(\xi)=P^-_0(\xi)$, both motor rows are in a stalled configuration: each of them exerts a non-zero opposing force, with net active force equal to zero \[ 0 = - \rho N \int_0^{\ell} P^+_0(\xi)\partial_{\xi}\Delta W(\xi)d \xi  + \rho N \int_0^{\ell}P^-_0(\xi)\partial_{\xi}\Delta W(\xi) d \xi.\]

Instead, in the one-row model, at steady-state equilibrium, the stalled motors exert a non-zero net force: \[ - \rho N \int_0^{\ell} P_0(\xi)\partial_{\xi}\Delta W(\xi) d \xi,\] where $P_0(\xi)$ is the equilibrium probability density of bound motors. This force causes filament displacement based on its stiffness $K$, breaking the symmetry of the ideal axoneme, where all microtubule doublets play identical roles. To restore symmetry, $\Delta W$ and $P_0$ can be chosen such that the net force at equilibrium is zero, as it was done in \cite{camalet2000generic}.

Exploiting standard assumptions \cite{camalet2000generic,guerin2011dynamical,guerin2011bidirectional}, we chose the potentials and the transition rates as
\begin{equation} \label{eq: two state model semplification}
\begin{array}{l}
    \Delta W(\xi) = U \cos (2 \pi \xi / \ell),\\ 
    \omega_1(\xi)+\omega_2(\xi) = \alpha,\\
    \omega_2(\xi) = \alpha\left(\eta + \frac{\Omega}{2 \pi^2} \left(\cos (2 \pi \xi/\ell) + \sin (2 \pi \xi/\ell) \right)\right),
\end{array}
\end{equation}
where $\alpha$ is a rate constants, $\eta$ is the average of the fraction of bounded motors
\begin{equation} \label{eta}
    \eta = \frac{1}{\ell}\int_0^{\ell}\frac{\omega_2(\xi)}{\alpha}\,d\xi,
\end{equation} 
 $\Omega$ represents the ATP concentration and $U$ is the potential energy in a tug-of-war unit. All these coefficients are $\xi$-independent. Notably, the sum of the transition rates is uniform as in \cite{camalet2000generic} but they are not symmetric, in the sense of \cite{guerin2011dynamical}, so that the $\mu$-chemoEH can effectively symmetrize the system composed of the two filaments and the motors.

We expand the probability densities using the Fourier series in space,
\begin{equation}\label{eq: P as Fourier series}
\begin{array}{ll}
     P^+(\xi,t) = \displaystyle p_0(t) + \sum\limits_{n>0} \left( p_n^+(t) \cos{\frac{2n \pi \xi}{\ell}} + q_n^+(t)\sin{\frac{2n \pi \xi}{\ell}}\right), \\
   P^-(\xi,t) = \displaystyle p_0(t) + \sum\limits_{n>0} \left( p_n^-(t) \cos{\frac{2n \pi \xi}{\ell}} + q_n^-(t)\sin{\frac{2n \pi \xi}{\ell}}\right).
\end{array}
\end{equation}
If we plug \eqref{eq: P as Fourier series} into the PDE system \eqref{eq:PDE motors} we get an infinite number of ODE systems for the Fourier coefficients of $P^\pm$. We observe that the coupling between the probabilities dynamics and the force balance equation \eqref{eq: force balance equation + filament} takes place only for the first order coefficients, $n=1$. This is therefore the only ODE that plays a role in the feedback between the motors' internal activity and the mechanical description of the flagella (see~\ref{appendix: tworow-higher order}).

For $n=1$, we call $p^\pm=p_1^{\pm}$ and $q^\pm=q_1^{\pm}$ for simplicity and obtain
\begin{equation} \label{eq: Fourier ODE first order p,q,f}
    \left\{\begin{array}{l}
\dot{p}^+(t) + \frac{2 \pi}{\ell} \dot{u}(t) q^+(t)= - \alpha p^+(t) + \alpha \Omega / (2 \pi^2 \ell),\\\\
\dot{q}^+(t) - \frac{2 \pi}{\ell} \dot{u}(t) p^+(t)= - \alpha q^+(t) + \alpha \Omega / (2 \pi^2 \ell),\\\\
\dot{p}^-(t) - \frac{2 \pi}{\ell} \dot{u}(t) q^-(t)= - \alpha p^-(t) + \alpha \Omega / (2 \pi^2 \ell),\\\\
\dot{q}^-(t) + \frac{2 \pi}{\ell} \dot{u}(t) p^-(t)= - \alpha q^-(t) + \alpha \Omega / (2 \pi^2 \ell),\\\\
 f(t) = -2\lambda \dot{u}(t)  - 2 K u(t) +  \rho N \pi U \left(q^+(t) - q^-(t)\right).
\end{array}\right.
\end{equation}

In this case, system \eqref{eq: microscopic system} reads $\underbar X = (p^+,q^+,p^-,q^-)^T$,
\[F( \underbar X,\dot u)= - \alpha  \underbar X+ \frac{2 \pi}{\ell}\begin{pmatrix} 
0 & - \dot u & 0 & 0\\
\dot u & 0 &0&0\\
0&0 & 0& \dot u \\
0&0&-\dot u & 0
\end{pmatrix}
X +  \begin{pmatrix}  \alpha \Omega / (2 \pi^2 \ell) \\ \alpha \Omega / (2 \pi^2 \ell)\\\alpha \Omega / (2 \pi^2 \ell)\\\alpha \Omega / (2 \pi^2 \ell)\end{pmatrix}, \] $G(X)= \rho \pi U( \underbar X \cdot \underbar {\text{e}}_2- \underbar X \cdot \underbar {\text{e}}_4)$ and $H(u,\dot u)=-2\lambda \dot{u}  - 2K u  $.

\subsection{The cubic model: an approximation} 
From the $\mu$-chemoEH, we derive a differential equation that relates the active force $f_a$ to the sliding velocity $\dot {u}$, embedding the motor dynamics probabilities into the resulting dynamics. This equation is referred to as the cubic model. The derivation partially follows \cite{julicher1997modeling} and the result is similar to the ODEs used in \cite{sartori2016dynamic}.
Differentiating the active force $f_a = - \rho N \pi U ( q^+ - q^-)$ with respect to time and substituting $\dot q^+$ and $ \dot q^-$ using the second and fourth equations of system \eqref{eq: Fourier ODE first order p,q,f}, we derive
\begin{equation}\label{eq: fadot}
    \dot{f_a} = - \alpha f_a + \rho N \pi U \frac{2 \pi}{\ell} \dot u (p^++p^-).
\end{equation}

We focus on the first four equations of system \eqref{eq: Fourier ODE first order p,q,f} and assume a small perturbation around the steady state. We then have $\dot u = \eps v$, with $|v| \leq 1$, and we expand $p^\pm$ and $q^\pm$ as follows
\begin{equation*}
   p^\pm = \sum_{n \geq 0} \tilde p^\pm_n \eps^n, \quad  q^\pm = \sum_{n \geq 0} \tilde q^\pm_n \eps^n,
\end{equation*}
where $\tilde p^\pm$ and $\tilde q^\pm$ satisfy
\begin{equation}  \label{1}
    \left\{\begin{array}{l}
0 = - \alpha \tilde p_n^\pm \mp \frac{2 \pi}{\ell} v \tilde q_{n-1}^\pm,\\
0 =  - \alpha \tilde q_n^\pm \pm \frac{2 \pi}{\ell} v\tilde p_{n-1}^\pm,
\end{array}\right.
\end{equation}
for $n \geq 1$, and $\tilde p_0^\pm = \tilde q_0^\pm= \Omega / (2 \pi^2 \ell)$.

Here, we assume that the rate of change of $v$ is small compared to $\alpha$. As a result, the equations that determine the probabilities coefficients can be treated as quasi-static, with all variables adapting to $v$ over time. In this context, setting the left-hand side of \eqref{1} to zero is reasonable, as the solution quickly converges to an equilibrium point determined by solving \eqref{1}. From equations \eqref{1}, we recover the recursion relations
\begin{equation*}
    \tilde p_n ^ \pm = \mp \frac{2 \pi}{\alpha \ell} v q_{n-1}^\pm, \quad \tilde q_n ^ \pm = \pm \frac{2 \pi}{\alpha \ell} v p_{n-1}^\pm.
\end{equation*}

Substituting these into equation \eqref{eq: fadot}, and retaining terms up to order three in $\eps$, we obtain the equilibrium relation
\begin{equation}\label{eq:equilib}
    0 = - \alpha f_a + 2 \rho N \Omega k_{cb} \eps v \left(1- \left(\frac{\eps v}{\gamma} \right)^2\right) + O(\eps^3),
\end{equation}
where $\gamma= \frac{\ell \alpha}{2 \pi}$ and $k_{cb}= U/ \ell^2$. Thus, we approximate equation \eqref{eq: fadot} by allowing the active force to evolve such that equation \eqref{eq:equilib} holds at equilibrium. The idea behind this procedure is to simplify system \eqref{eq: Fourier ODE first order p,q,f} while preserving its linear characteristic. The coefficient $\gamma$ is determined by the system's chemical properties, rather than selected \textit{a posteriori}, as in \cite{sartori2016dynamic}. 

System \eqref{eq: microscopic system} is now defined by imposing $X = f_a$ 
\begin{equation}\label{26112401}
    F(X,\dot u)= - \alpha f_a+ 2 \rho N k_{cb} \Omega \dot u(1 - \frac{ \dot u^2}{\gamma^2}),
\end{equation}
and $H(u, \dot u) = -2\lambda \dot u - 2K u$.
Notice that, in the frequency domain, equation \eqref{26112401} becomes the same as the non linear relation used in \cite{hilfinger2009nonlinear}.
\subsection{The chemoEH model}
We present the third and last model, the chemoEH model \cite{oriola2017nonlinear}, based on the qualitative description made in \cite{riedel2007molecular}. The set up is the same as the one explained for the $\mu$-chemoEH, where one fixes an arc-length $s \in [0,L]$ and considers a tug-of-war unit with two opposing filaments of length $\ell$ as shown in Figure~\ref{fig:axoneme}(C). In a tug-of-war unit, $N$ is the total number of motors and $n^{\pm}(s,t)$ is the number of motors that are anchored to the $\pm$ filament and bound (state 1) to the opposite filament. Let us define $F^{\pm}(s,t)$ as the load per motor each group of motors experiences caused by the action of the antagonistic group. The load $F^{\pm}$ satisfies a linear force-velocity relationship where $F^{\pm}=F^{\pm}(\dot u)= \pm f_0 (1 \mp \dot u/v_0)$. The stall force $f_0$ is defined as the absolute value of the load a motor experiences at stall, while $v_0$ is the relative velocity between the filaments at zero load. The external force is then balanced by an active part and a resistance to the sliding
\begin{equation}\label{eq: Gadelha f}
\begin{split}
    f(s,t) &= \rho (n^+ F^+ + n^-F^-)-K u \\
    & = \rho f_0 \left((n^+-n^-) - (n^+ + n^-)  \frac{\dot u}{v_0} \right)-K u .
\end{split}
\end{equation}
The evolution law for the quantities $n^{\pm}$ reads
\begin{equation}\label{eq: Gadelha nplus, nminus}
    \dot n^{\pm}= \pi_0(N-n^{\pm})- \eps_0 n^{\pm} \exp(\pm F^{\pm}/f_c),
\end{equation}
where $\pi_0$ and $\eps_0$ are constants rates and $f_c$ is the characteristic unbinding force. In this model, the ratio $\bar f = f_0 / f_c$ is held constant, while $f_0$ varies, reflecting the potential energy available to the motors, which is determined by the amount of ATP present in the system. In this case, system \eqref{eq: microscopic system} reads $\underbar X = (n^+,n^-)^T$,
\[F( \underbar X,\dot u)= - \pi_0  \underbar X+ \eps_0 e^{\bar f}\begin{pmatrix} 
 e^{-\bar f\frac{\dot u}{v_0}} &  0\\
0 &  e^{\bar f\frac{\dot u}{v_0}} \\
\end{pmatrix}
\underbar X +  \begin{pmatrix}  
\pi_0 N \\
\pi_0 N
\end{pmatrix}, \] $G(\underbar X,\dot u)= \rho (F^+,F^-)^T \cdot \underbar X$ and $H(u)=-K u  $.

\subsubsection{Comparison between $\mu$-chemoEH and chemoEH model}

In the following, we 

highlight the main differences between the $\mu$-chemoEH model and the chemoEH model, as they both aim to represent how the internal two-state chemistry of motors results in a sliding feedback mechanism.

Since the chemoEH model does not include the microscopic space variable $\xi$, it could be qualitatively interpreted as describing the average dynamics of the $\mu$-chemoEH model over the tug-of-war unit. However, mathematical analysis of the two models demonstrates that the density of bound motors $n^{\pm}(t)/(N \ell)$ represents neither the average $p_0$ nor the first-order coefficients $p^{\pm}$ and $q^{\pm}$ of the Fourier expansion in \eqref{eq: P as Fourier series}.

This is due to key differences, particularly in the transition rates between states. In the $\mu$-chemoEH model, assumption \eqref{eq: two state model semplification} is such that the sum of the transition rates remains constant in both space and time, while in the chemoEH model this is generally not the case, except when $\dot u = 0$
and the sum is $\pi_0 + \eps_0 e^{\bar f}$. Indeed, the derivation of the chemoEH model follows from \cite{grill2005theory}, where the main assumptions are that the unbind rate is load dependent and therefore velocity dependent, and that the motors have a detachment rate which is higher than the attachment one. These hypotheses are not compatible with
the hypothesis of uniform sum of transition rates 

formulated in \cite{guerin2011dynamical}, which holds when the two rates are almost uniform. Finally, in the chemoEH model there is no passive viscosity at play, $\lambda =0$, but only an active viscosity, given by

the term $(n^+ + n^-)  \dot u / v_0$.

A comparison between the order of magnitude of the parameters appearing in the two models is discussed in \cite{JulicherForceGenerator}, based on a specific choice for $\Delta W$. We propose a similar approach, using the difference of potentials defined in \eqref{eq: two state model semplification}. We focus on the forces exerted on on the upper filament: the force $f^+$ in equation \eqref{eq:fplus} for the $\mu$-chemoEH and the force $f_{cEH}^+= \rho n^+ F^+$. When the system \eqref{eq: Fourier ODE first order p,q,f} reaches the equilibrium, the force in equation \eqref{eq:fplus} becomes $ f^+=\rho N U \Omega / (2 \pi \ell) $. In the chemoEH model, the corresponding equilibrium force is $ f_{cEH}^+=n_{eq} \rho f_0$, where $n_{eq}= N \pi_0 / (\pi_0 + \eps_0 e^{\bar f})$, as defined in~\ref{appendix: chemoEH}. It follows that, to achieve comparable force magnitudes in the two models, the following relation must hold 
\begin{equation} \label{magnitudeorder}
    f_0 \sim \Omega \frac{U}{2 \pi \ell}\frac{n_{eq}}{N}.
\end{equation}

Furthermore, when the two models describe the isolated axoneme, with $f=0$, the linear analysis of the two ODE systems \eqref{eq_isolated: Fourier ODE first order p,q,f=0} and \eqref{eq: ode n+n-f=0} reveal that the role of $n^+-n^-$ is equivalent to the one of $q^+-q^-$, implying a strict correlation between the two systems. Instead, the two models differ in their non linear part, as observed in~\ref{appendix: chemoEH}. 

As noted in \cite{cass2023reaction}, in the chemoEH there are two characteristic times $\tau_0=(\pi_0+\eps_0)^{-1}$ and $n_0/\pi_0=(\pi_0+\eps_0 e^{\bar{f}})^{-1}$. We chose the first one to be related to the characteristic dynein switching rate in the $\mu$-chemoEH, by imposing $\alpha = \tau_0^{-1}$.

\section{Linear stability analysis} \label{sec: linear study}
In this section we perform the linear stability analysis of complete active axoneme by taking the $\mu$-chemoEH \eqref{eq: Fourier ODE first order p,q,f} as the form of activation to be coupled with the scalar moment balance equation \eqref{eq: local moment balance}. Notice that the analysis would have been the same with the cubic model, which is introduced as an approximation of the $\mu$-chemoEH non linearities.

The linearization of \eqref{eq: local moment balance} around the equilibrium $\varphi=0$ is
\begin{equation} \label{eq: linearized beam}
    \xi_n \dot{\varphi}(s,t) + B \varphi''''(s,t) +a f''(s,t)   =0.
\end{equation}
This equation is obtained by differentiating twice equation \eqref{eq: local moment balance}, using the relation $(\dot{\mathbf{r}})' = \dot{\varphi}\mathbf{n}$, then applying the small deformation approximation to remove nonlinear terms. Assuming a linear perturbation of the form $f(s,t)=\tilde f(s)e^{\sigma t}$ and $\varphi(s,t)=\tilde \varphi(s)e^{\sigma t}$, where $\sigma = \mu + i\theta$, equation \eqref{eq: linearized beam} becomes a fourth-order ODE for $\tilde \varphi(s)$
\begin{equation} \label{eq: linearized beam with f}
    \xi_n \sigma \tilde{\varphi}(s) + B \tilde \varphi''''(s) - a^2 \chi(\Omega, \sigma) \tilde \varphi ''(s)=0.
\end{equation}

The coefficient
\begin{equation}\label{lin. coeff}
    \chi(\Omega, \sigma)= 2\left(  \lambda \sigma+  K - \rho N k_{cb} \Omega \frac{\sigma}{(\alpha + \sigma)} \right),
\end{equation}
with $k_{cb}= U/ \ell^2$, is recovered from system \eqref{eq: Fourier ODE first order p,q,f} and relates the force $f$ and the sliding $u$ at the linear level in such a way that $\tilde f = -  \chi(\Omega, \sigma) \tilde u$.
Notice that the linear response coefficient \eqref{lin. coeff} is twice 
the one defined in \cite{camalet2000generic}.

We bring the system to a non-dimensional form by defining $\bar{s} = s/L$ and $\bar{t}= t \alpha$. The non-dimensional linear response coefficient becomes $\bar{\chi} =  a^2 L^2 \chi/B$ and the ratio between viscous and elastic forces, called sperm number Sp, is defined as $\text{Sp} = L (\xi_n \alpha/B)^{1/4}$. The non dimensional system of equation \eqref{eq: linearized beam with f} together with clamped-free boundary conditions \eqref{BC} becomes
\begin{equation}\label{eq: boundary value problem Fourier space}
  \begin{split}
     &  \bar \sigma \text{Sp}^4  \varphi  + \varphi'''' - \bar{\chi}  \varphi'' = 0,\\
     &\varphi(0) = 0,\quad \varphi'''(0) - \bar{\chi} \varphi'(0)=0,\\
     &\varphi''(1) - \bar{\chi}\varphi(1)=0, \quad \varphi'(1)=0,
     \end{split}
\end{equation}
 where we dropped the $\tilde{}$ symbol on $\varphi$.
Since the linearization of \eqref{eq: local moment balance} implies that the tension is null, we only have four boundary conditions instead of six. The third equation of system \eqref{eq: boundary value problem Fourier space} is derived from having zero velocity at the base, and hence $T'(0)=0$.

The general solution to system \eqref{eq: boundary value problem Fourier space} is $\varphi(s) = \sum_{j = 1}^4A_je^{\beta_j\bar s}$, where $\beta_j$ are the roots of the characteristic equation $\beta_j^4 - \bar{\chi}\beta_j^2 + \bar{\zeta}=0,$ while $A_j$ are the amplitudes determined by the boundary conditions.

The boundary conditions form a homogeneous system which admits non trivial solutions for the amplitudes $A_j$ only if its determinant, defined as $\Lambda = \Lambda(\bar{\chi},\bar{\sigma};\Omega)$, is null.

For a fixed $\Omega$, the equation $\Lambda(\bar{\sigma}; \Omega) = 0$ yields multiple solutions $\bar{\sigma}_n$, ordered by their real parts in such a way that $ \mu_n> \mu_{n+1} $. We refer to these solutions as modes. This ensures that the first eigenvalue to cross the imaginary axis is $\bar{\sigma}_1$. At the critical ATP concentration $\Omega = \Omega_c$, defined by $\mu_1(\Omega_c) = 0$ and $\theta_c = \theta_1(\Omega_c)$, the system is purely oscillatory.

\subsection{Small Machin number}
Consider an oscillatory ansatz $\varphi(s,t)= \tilde{\varphi}(s) e^{i \theta t}$ for the solution of equation \eqref{eq: linearized beam with f}. One can introduce an important non-dimensional quantity, the Machin number \cite{machin1958wave}, defined as \begin{equation} \label{mach}
    \text{Ma} = \frac{L}{\ell_{\text{Ma}}},\quad \ell_{\text{Ma}}=\left(\frac{B}{\theta \xi_n}\right)^{1/4},
\end{equation} 
with ${\ell_{\text{Ma}}}$ being the characteristic length of the system. When the system 

oscillates with frequency $\theta$, the Machin number is therefore such that $M_a=\bar \theta^{\frac{1}{4}} \text{Sp}$, where $\bar \theta$ is the dimensionless frequency. The Machin number is a natural parameter for studying systems where the oscillation frequency is known a-priori. In contrast, the sperm number is commonly used when the oscillation frequency emerges from an instability, and is an unknown of the problem.

If we consider small flagella such that $L \ll {\ell_{\text{Ma}}}$, we can study the limit ${\text{Ma}} \to 0$, which in turn implies $\theta \to 0$. In this case, the linear response coefficient reduces to the non-dimensional elastic term $\chi(\Omega, i \theta) \to K$. The non-dimensional linear equation for $\varphi(s)$ becomes $\varphi'''' - \mu_e  \varphi'' = 0$, whose solution is a standing wave, suggesting that small Machin number flagella are not well represented by a sliding feedback model. 

A typical small Machin number flagellum is the \textit{Chlamydomonas} ($L \approx 10 \,\mu m$), where experimental data in \cite{sartori2016dynamic} reveal a critical frequency of $\theta_c = 2\pi \cdot 40 \, \text{s}^{-1}$ and therefore an estimated Machin number ${\text{Ma}}\approx 2 $, which the authors considered sufficiently small, in comparison with the bull sperm case ($L \approx 50\, \mu m$), where the ratio is instead 
${\text{Ma}} \approx 8$.

In \cite{sartori2016dynamic}, the sliding feedback model did not match well the experimental data for \textit{Chlamydomonas} flagella. The authors attributed this discrepancy to the small ratio $L/{\ell_{\text{Ma}}}$ and the reasoning above. However a recent study \cite{cass2023reaction}, which developed an alternative non-linear sliding feedback model, showed a better fit with the experimental data. This could be explained by the fact that non-linearities are crucial to the system and/or that the Machin number is not small enough to justify the assumption.
For this reason, we find useful to perform simulation of the $\mu$-chemoEH sliding-feedback model also on short flagella.

\section{The case of long flagella: numerical study}
\subsection{Near the bifurcation}
In this chapter we present the case study of bull sperm, commonly proposed as a prototype for long flagella \cite{oriola2017nonlinear,bayly2015analysis}. We use the $\mu$-chemoEH model with the parameters outlined in Table~\ref{tab:ParametersBullVSChl}. 
\begin{figure}[htb!]
    \centering
    \begin{tabular}{cc}
        {\subfloat[]{\includegraphics[width=0.4\linewidth]{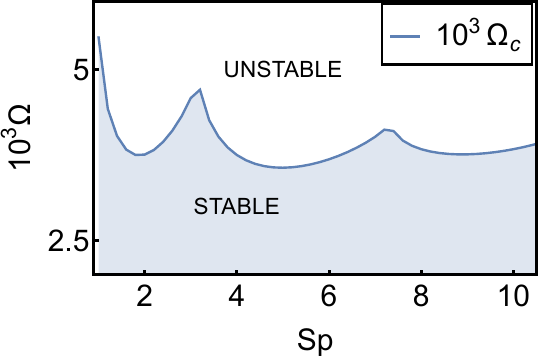}}}&
        \subfloat[]{\includegraphics[width=0.4\linewidth]{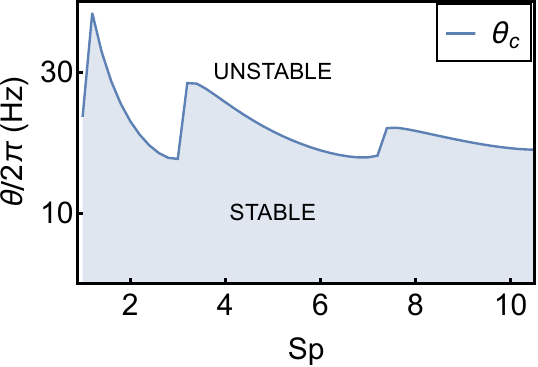}} \\
          \subfloat[]{\includegraphics[width=0.4\linewidth]{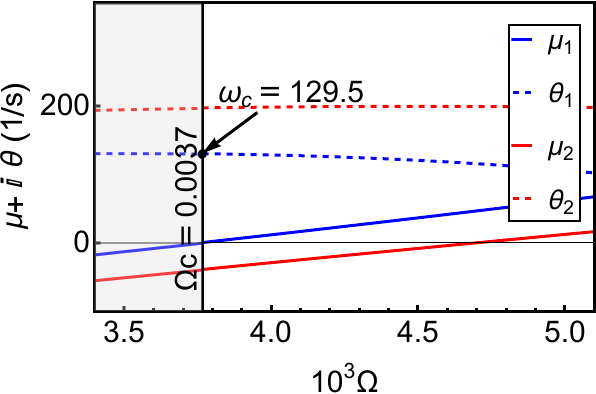}} &
       \hspace{10mm}\raisebox{3mm}{\subfloat[]{\includegraphics[width=0.4\linewidth]{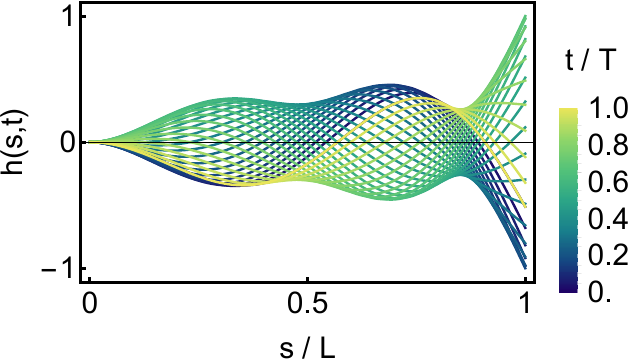}} }\\
    \end{tabular}
\caption{Linear study for the two row model.  (a), (b) Critical ATP concentration $10^3\Omega_c$ and critical frequency $\theta_c$ varying Sp. (c) Real (solid lines) and imaginary (dots, units are rad/s) parts of the first two modes varying the activation parameter $10^3\Omega$. (d) Transverse deformation $h(\bar s)$ over one beating cycle.}
\label{fig: LinearStudyBullSperm}
\end{figure}
To facilitate the comparison with the literature we simulate the $\mu$-chemoEH model by halving the motor force $f(t)$ defined in \eqref{eq: Fourier ODE first order p,q,f}; in this way we obtain that the linear analysis matches the one in \cite{camalet2000generic}.
\begin{table}[ht!]
    \centering
    \begin{tabular}{|c|c|c|c|c|}
    \hline
        Param. & Definition & Units & Value & Ref. \\
        \hline
        $L$ & Flagellum length & $\mu$m & 58.3 & \cite{bayly2015analysis} \\
        $B$ & Bending rigidity & $\text{pN} \cdot \mu \text{m}^2$ & 1700 & \cite{bayly2015analysis}\\
        $a$ & Inter-filament distance & $\mu$m & 0.185 & \cite{bayly2015analysis} \\
        $\xi_n$ & Normal RFT coefficient & $\text{pN} \cdot \text{s} / \mu \text{m}^2$ & 0.0034 & \cite{bayly2015analysis} \\
        $K$ & Internal elasticity & $\text{pN} / \mu \text{m}^2$ & 228 & \cite{sartori2016dynamic} \\
        $\lambda$ & Internal viscosity & $\text{pN} \cdot \text{s} / \mu \text{m}^2$ & 7 & \\
        $\alpha$ & Dynein rate constant & 1/s & 250 & \cite{bayly2015analysis} \\
        $\ell$ & Length of a tug-of-war cell & $\mu \text{m}$ & 1 & \cite{oriola2017nonlinear} \\
        $\rho$ & Density of tug-of-war cells & $1/\mu \text{m}$ & 1 & \cite{oriola2017nonlinear} \\
        $N$ & Motors in a tug-of-war cell & 1 & $10^3$ & \cite{oriola2017nonlinear}\\
        $k_{cb}$ & Motor domain's stiffness & $\text{pN}/\mu\text{m}$ & $10^3$ & \cite{camalet2000generic}\\
        \hline
    \end{tabular}
    \caption{Bull sperm parameters}
    \label{tab:ParametersBullVSChl}
\end{table}

The objective of the following section is twofold. First, we aim to confirm the agreement between the linear study and the fully non-linear simulations for the $\mu$-chemoEH model. Secondly, we investigate the coordination of molecular motors' probabilities appearing in equation \eqref{eq: Fourier ODE first order p,q,f}. 

The linear analysis for the two row model is carried out as presented in Section~\ref{sec: linear study}. To reproduce a beating frequency close to $20 \cdot 2 \pi$ rad/s -- which is the one shown in \cite{riedel2007molecular} -- we adjust the internal friction $\lambda$ to 7 $\text{pN} \cdot \text{s} / \mu \text{m}^2$ (which is comparable to the one from \cite{camalet2000generic}). In Figure~\ref{fig: LinearStudyBullSperm}(c) we observe the behavior of the real and imaginary part of the first two modes $\sigma_j=\mu_j+i \theta_j$ with $j=1,2$ as functions of the activation parameter $\Omega$ . The first mode crosses the imaginary axis with a critical frequency $\theta_c = 20.6$Hz and at $\Omega_c \simeq 3.7 \times 10^{-3}$. Below this critical value all the modes are stable. Knowing the critical frequency, one can compute the Machin number \eqref{mach} $\text{Ma}= 7.4$, which in this case indicates that the system is in a high Machin number regime, where the sliding control model fits well the experimental data. Instead, the sperm number is $\text{Sp} = 8.71$.

In Figures~\ref{fig: LinearStudyBullSperm}(a) and (b), we depict how the activation parameter $\Omega_c$ and the critical frequency $\theta_c$ respond to changes in the dimensionless ratio Sp. The stable region, in blue, corresponds to a non-oscillating system. These results align with the ones presented in \cite{bayly2015analysis}. In practice, each value of the parameter Sp corresponds to a simulation with a distinct flagellum length $L$, determined by the formula $L = \text{Sp} \cdot (\alpha \xi_n / B)^{-1/4}$; this length varies between $L \sim 6 \mu \text{m}$ and $L\sim 60 \mu \text{m}$. Finally, in Figure~\ref{fig: LinearStudyBullSperm}(d) we observe the behavior of the transverse deformation $\mathbf{r}(\bar s,\bar t)\cdot \mathbf{e}_2$ in the small amplitude regime, defined as
\[h(\bar s)= \int_0^{\bar s}\frac{\varphi( y,\bar t)}{|\varphi( y,\bar t)|} \, dy.\]
The angle $\varphi$ is the solution of \eqref{eq: boundary value problem Fourier space} corresponding to the critical bifurcation parameters $(\sigma,\Omega)=(i \theta_c,\Omega_c)$.
\begin{figure}[hbt!]
\centering
    \subfloat[]{
    \includegraphics[width=0.45\linewidth]{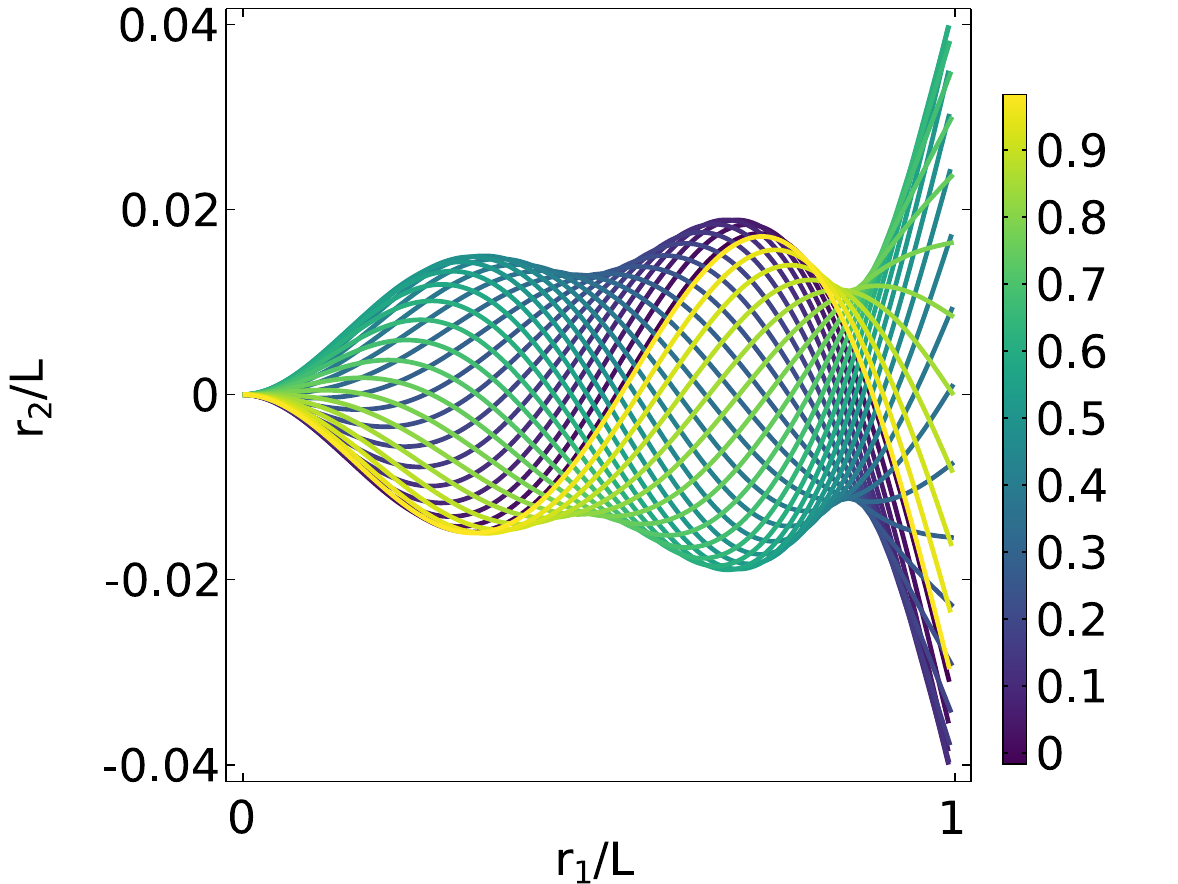}}
    \hfill
    \subfloat[]{\includegraphics[width=0.45\linewidth]{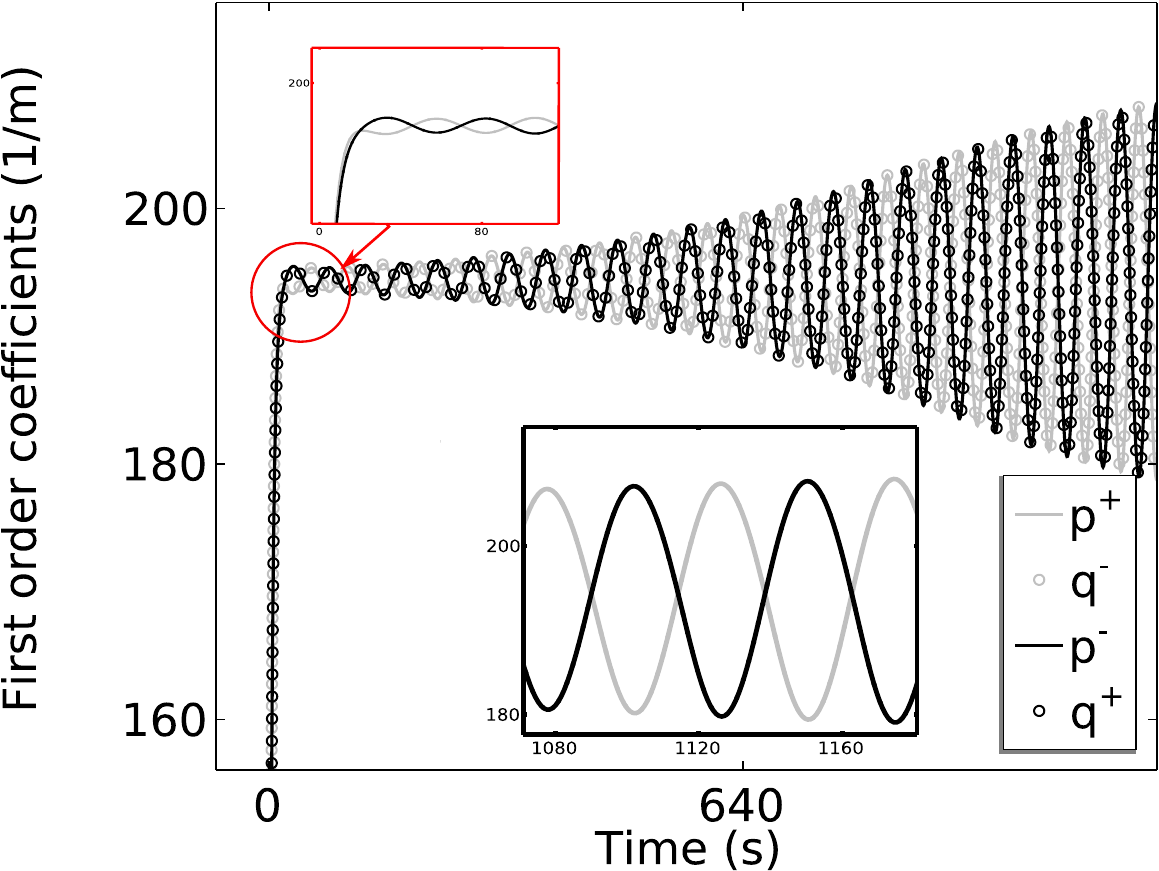}}
    \caption{Fully non-linear simulations for the $\mu$-chemoEH model with $\eps=0.01$. (a) Deformed configuration during a beating cycle. The legend bar on the right indicates the time normalized over the period of oscillation $t/T$. (b) Oscillations in time of the first order for the probabilities $p^\pm(t)$ (solid gray and black line) and $q^\pm(t)$ (dotted black and gray line). The inlets are a close-up of the transient interval of time (in red) and one of the limit cycle regime (in black) where we only show $p^{\pm}$.}
    \label{fig: Tworow eps=0}
\end{figure}

We now investigate the behavior of the fully non-linear coupled system of equations given by \eqref{eq: local moment balance} and \eqref{eq: Fourier ODE first order p,q,f}, under clamped-free boundary conditions and in the time domain.
Numerical simulations are conducted using COMSOL software (see ~\ref{app:numerics}). We define the parameter $\eps = |\Omega - \Omega_c| / \Omega_c$ to represent the relative deviation of the ATP concentration $\Omega$ from the critical bifurcation value $\Omega_c$, determined through the linear analysis. To observe the bifurcated oscillatory solution, we apply an initial condition for the probabilities that slightly deviates from equilibrium:
\[p^{\pm}(0,s)=\frac{\beta}{\ell}(1\pm0.01), \quad  q^{\pm}(0,s)=\frac{\beta}{\ell}(1 \mp 0.01).\]
In Figure~\ref{fig: Tworow eps=0} we utilize the parameters in Table~\ref{tab:ParametersBullVSChl} to run simulations at the instability threshold, where $\eps = 0.01$ and spontaneous oscillations take place. The deformed configuration of the flagellum during a beating cycle, defined as the curve $(r_1,r_2) =(\mathbf{r}\cdot \mathbf{e}_1,\mathbf{r}\cdot \mathbf{e}_2)$, is shown in Figure~\ref{fig: Tworow eps=0}(a) and closely matches the one of Figure~\ref{fig: LinearStudyBullSperm}(d). The non-linear simulations reproduce correctly the linear predictions of the critical activation $\Omega_c$ and of the frequency of oscillation $\theta_c$. 

Moreover, once the critical ATP threshold is reached, the first-order Fourier coefficients $ p^\pm(t) $ and $ q^\pm(t) $ of the probabilities \eqref{eq: P as Fourier series} begin to oscillate in synchrony, as shown in Figure~\ref{fig: Tworow eps=0}(b). Specifically, we observe that $ p^+ = q^- $ and $ p^- = q^+ $, with $ p^+ $ and $ q^+ $ exhibiting a phase shift of half a period relative to each other. The complete expression for the $\xi$-dependent probabilities $ P^\pm(\xi, t) $ in Equation \eqref{eq: P as Fourier series} is given by\[
\begin{aligned}
     P^+(\xi, t) &=  p_0(t) +  p^+(t) \cos{\frac{2 \pi \xi}{\ell}} + q^+(t)\sin{\frac{2 \pi \xi}{\ell}}, \\
     P^-(\xi, t) &= p_0(t)  + q^+(t) \cos{\frac{2 \pi \xi}{\ell}} + p^+(t)\sin{\frac{2 \pi \xi}{\ell}}.
\end{aligned}
\]
Here, we find that $ P^-(\xi, t) = - P^+\left(\frac{3}{4} \ell - \xi, t\right) $, implying a spatial phase shift of three-quarters of a period.

\begin{figure}[!htb]
\minipage{0.32\textwidth}
  \includegraphics[width=\linewidth]{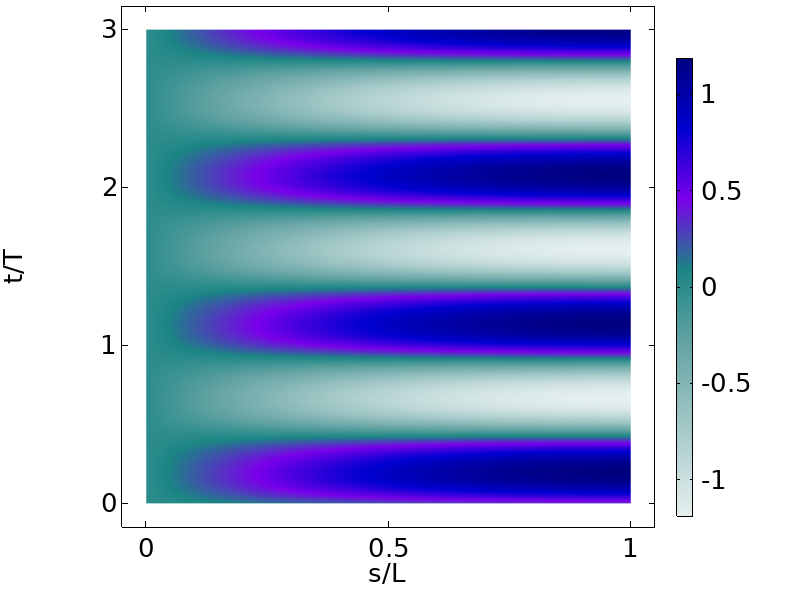}
  \caption*{Sp=1}
\endminipage\hfill
\minipage{0.32\textwidth}
  \includegraphics[width=\linewidth]{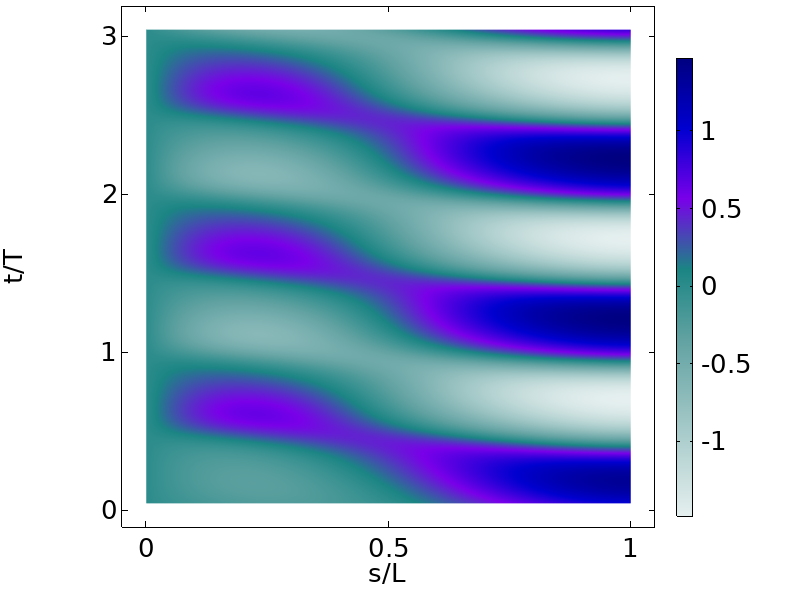}
    \caption*{Sp=5}
\endminipage\hfill
\minipage{0.32\textwidth}%
  \includegraphics[width=\linewidth]{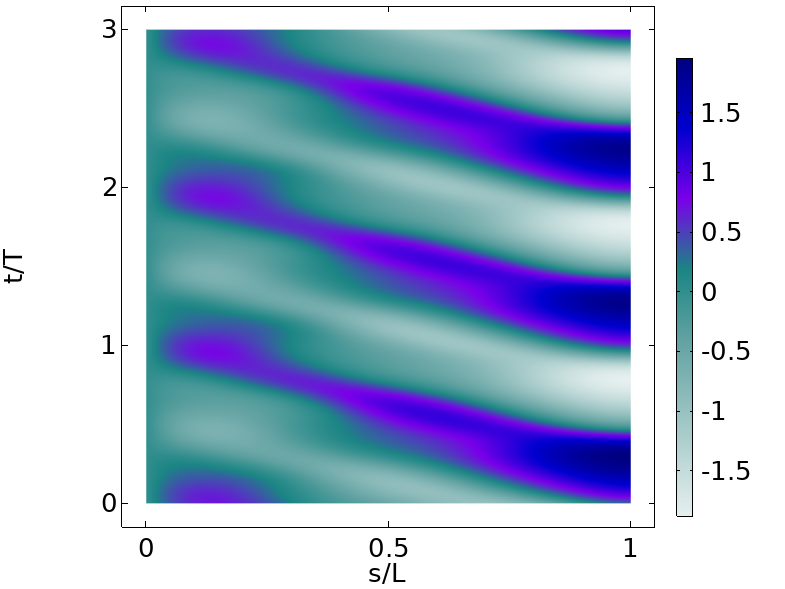}
    \caption*{Sp=10}
\endminipage
\caption{Kymographs of the tangent angle  $\varphi$ varying the sperm number Sp $\in \{1,5,10\}$. The legend bar on the right indicates the amplitude of $\varphi$. The tangent angle is plotted over the dimensionless arc-length and three period of oscillations. The wave speed increases by increasing Sp, going from a nearly standing wave to a traveling wave. All the parameters, except the length $L$, are defined in Table~\ref{tab:ParametersBullVSChl}.}
\label{spermchange}
\end{figure}

To conclude this section, we present non-linear simulations for the bull sperm with $\eps=0.5$, which is close enough to the bifurcation. We observe a typical known behavior in flagella \cite{machin1958wave}:
by increasing the parameter Sp,  the traveling wave velocity increases. In Figure~\ref{spermchange} the kymographs of the tangent angle $\varphi(s,t)$ are plotted against the non-dimensional flagellum length and over three periods of oscillations. The total length of the flagellum is varied by imposing different values of sperm number. When Sp=1 the viscous and bending forces are of comparable magnitude, and we observe almost standing waves: the flagellum does not generate any propulsion in the fluid. The more the viscous forces overcome the bending ones, the more we observe retrograde (tip to base) traveling waves with a significant speed.

\subsection{Away from the bifurcation: comparison with the cubic model}
\begin{figure}[ht!]
\centering
    \subfloat[]{
    \includegraphics[width=0.4\linewidth]{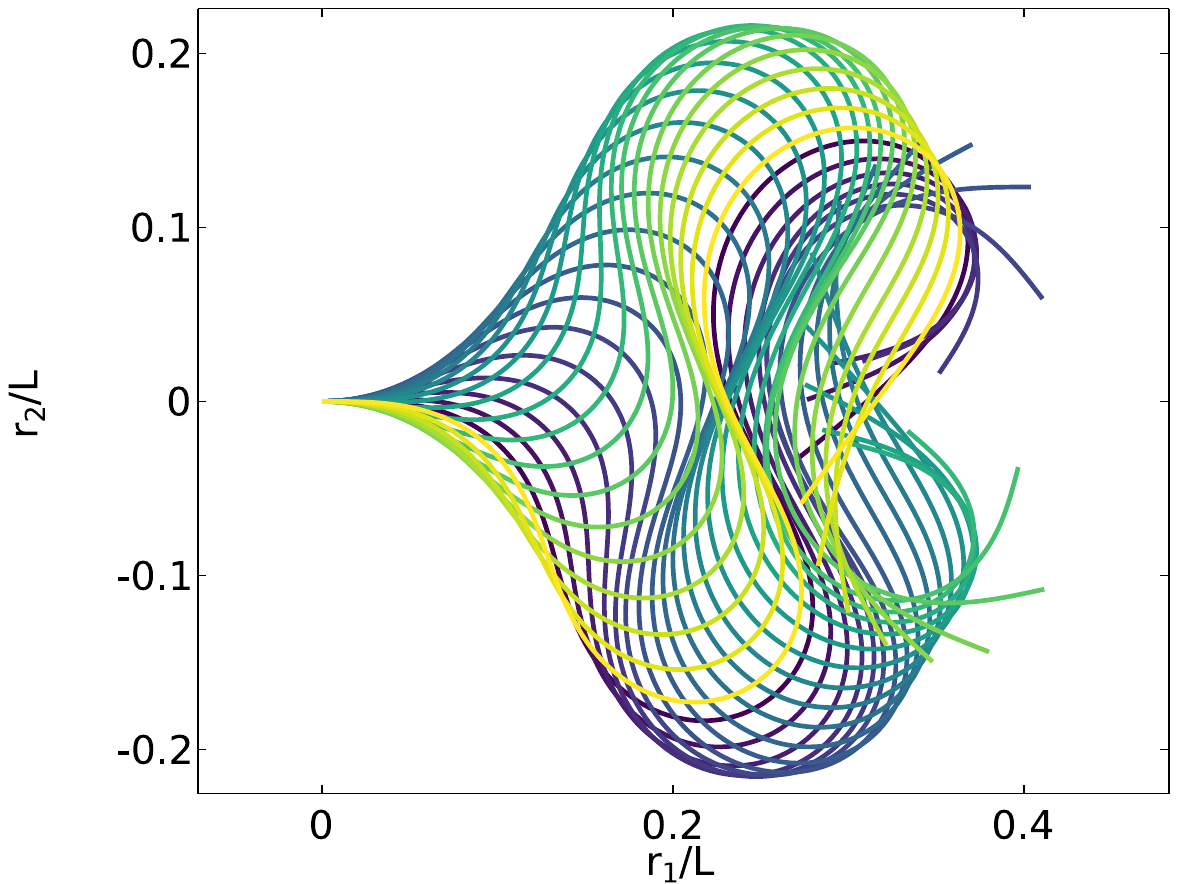}}
    \subfloat[]{\includegraphics[width=0.4\linewidth]{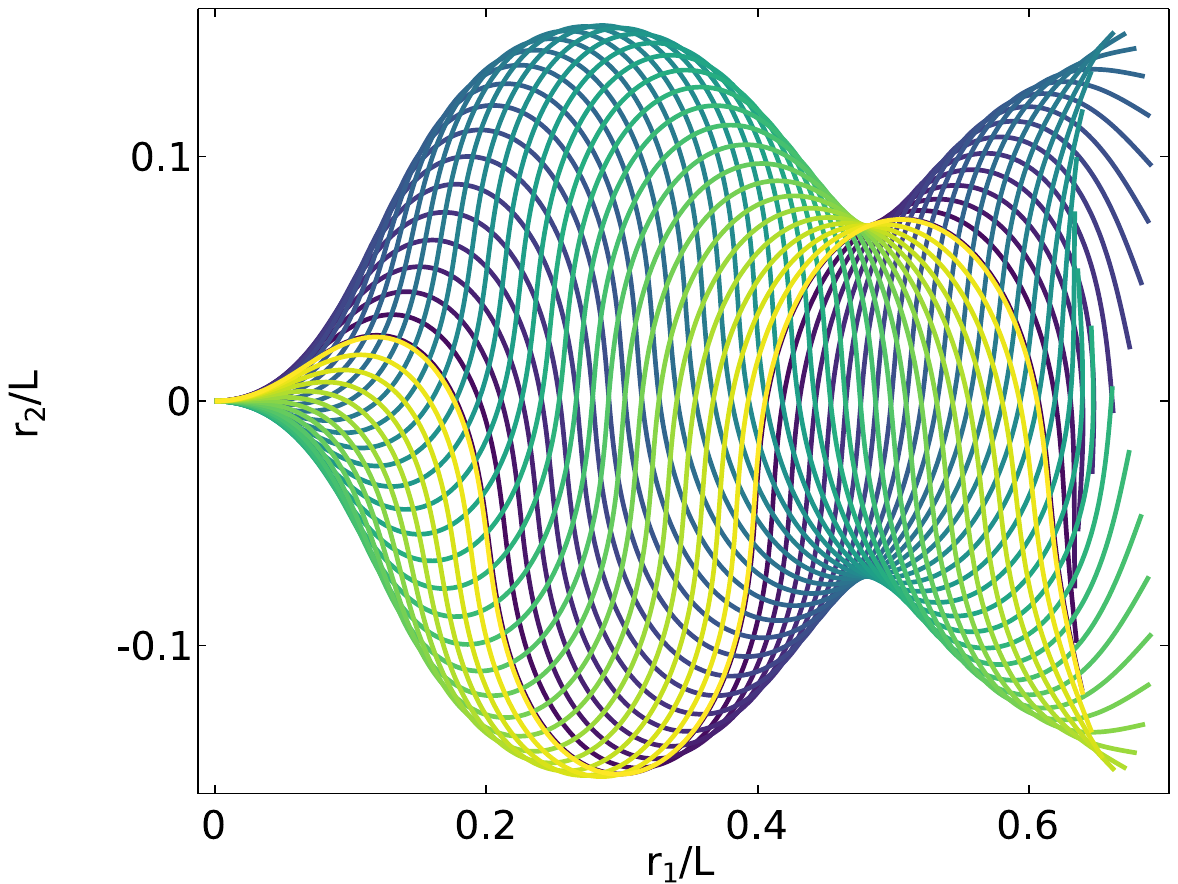}}
    \caption{Fully non-linear simulations for the deformed configuration of bull sperm over a beating cycle. (a) The $\mu$-chemoEH at $\eps=2$ and (b) the cubic model at $\eps=4$. Parameters in Table~\ref{tab:ParametersBullVSChl}}
    \label{fig: large amplitude comp}
\end{figure}

In this section, we present a
comparison between the $\mu$-chemoEH model (Figure \ref{fig: large amplitude comp}(a)) and the cubic model (Figure \ref{fig: large amplitude comp}(b)), in the regime of large-amplitude deformations.
We recall that the latter is derived from the former by means of an expansion analysis in the vicinity of the equilibrium point of system \eqref{eq: Fourier ODE first order p,q,f}. 

The $\mu$-chemoEH model exhibits a greater sensitivity to the $\eps$ parameter than the cubic model. Even for the same small value of $\eps$, the $\mu$-chemoEH flagellum shows a larger deviation from the straight equilibrium configuration compared to the cubic model. This trend holds even for larger values of $\eps$, which is why we present the waveform for $\eps = 2$ in the case of the $\mu$-chemoEH model and for $\eps = 4$ for the cubic model. As shown in Figure \ref{fig: large amplitude comp}, the distinction is apparent: in the $\mu$-chemoEH, the nonlinear terms are such that the flagellum exhibits more complex deformations with respect to the cubic model.

\section{The case of short flagella: numerical study}
\subsection{Comparison between the three models}

\begin{table}[ht!]
    \centering
    \begin{tabular}{|c|c|c|c|c|c|}
    \hline
        & Param. & Definition & Units & Value & Ref. \\
    \hline
        \multirow{9}{*}{{All}}
        & $L$ & Flagellum length & $\mu$m & 12 & \cite{sartori2016dynamic} \\
        & $B$ & Bending rigidity & $\text{pN}\cdot \mu \text{m}^2$ & 840 & \cite{cass2023reaction}\\
        & $a$ & Inter-filament distance & $\mu$m & 0.06 & \cite{sartori2016dynamic} \\
        & $\xi_n$ & Normal RFT coefficient & $\text{pN} \cdot \text{s} / \mu \text{m}^2$ & 0.0034 & \cite{sartori2016dynamic} \\
        & $K$ & Internal elasticity & $\text{pN} / \mu \text{m}^2$ & $25 \times 10^3$ & \\
        & $\alpha$ & Dynein rate constant & $\text{s}^-1$ & 200 & \\
        & $\ell$ & Length of a tug-of-war cell & $\mu \text{m}$ & 1 & \cite{oriola2017nonlinear} \\
        & $\rho$ & Density of tug-of-war cells & $\mu \text{m}^{-1}$ & 1 & \cite{oriola2017nonlinear} \\
        & $N$ & Motors in a tug-of-war cell & 1 & $10^3$ & \cite{oriola2017nonlinear} \\
    \hline
        \multirow{2}{*}{{$\mu$-chemoEH}}
        & $\lambda$ & Internal viscosity & $\text{pN} \cdot \text{s} / \mu \text{m}^2$ & 4 & \\
        & $k_{cb}$ & Motor domain stiffness & $\text{pN}/\mu\text{m}$ & $10^3$ & \cite{camalet2000generic} \\
    \hline
        \multirow{2}{*}{{chemoEH}}
                & $v_0$ & Velocity at zero load & $\mu\text{m}/\text{s}$ & 60 & \\
        & $\bar f$ & $f_c/f_0$ & 1 & 2 & \cite{oriola2017nonlinear}\\
        & $\eta$ & Average of bound motors & 1 & 0.5 & \\
    \hline
    \end{tabular}
    \caption{\textit{Chlamydomonas} parameters for the $\mu$-chemoEH/cubic model and for the chemoEH model}
    \label{tab:Chl}
\end{table}
In the following chapter we focus on the results of nonlinear simulations of \textit{Chlamydomonas}, using the parameters in Table~\ref{tab:Chl}. The parameters are chosen on the basis of previous works, with some adjustment to make the models more comparable. Recall that the dynein switching rate $\alpha$ is related with the characteristic time of the chemoEH model $\tau_0$ as $\alpha = \tau_0^{-1}$. The internal elastic constant $K$ is defined in such a way that $a^2 K = 90 \text{pN}$, and the velocity $v_0 = 60 \mu\text{m}/\text{s}$ is such that $\zeta= a/ (\tau_0 v_0)= 0.4$; both of these choices were made to be close to the range of parameters used for the short flagella in \cite{cass2023reaction}. In the two-row model, the parameter $\eta$ does not influence the computation of the active force $f_a$ \eqref{fatworow} because it only appears in the average $p_0$ of $P^\pm$. Since $p_0$ cancels out during the calculation of $f_a$, $\eta$ has no effect on coupled system. In contrast, in the chemoEH model, the rates $\pi_0$ and $\eps_0$ depend explicitly on $\eta$, making it an essential parameter in the formulation of the model. Once the parameter setting is defined, by gradually increasing the distance $\eps$ from the instability, we compare the results given by three different non-linear model for flagellar beating activation: the $\mu$-chemoEH, the chemoEH model and the cubic model. 

The linear analysis gives the bifurcation values: for the $\mu$-chemoEH and cubic model the onset of oscillation is at $\Omega_c=0.0384$, while for the chemoEH we have $f_0=10.9\text{pN}$. Therefore, the term given on the right side of the relation \eqref{magnitudeorder} is $51.3 \text{pN}$, which makes the order of magnitude comparable. 

We start close to the instability, with $\eps=0.01$, and then grow further away from it. In all of the following cases we analyze the system once it has entered in its limit cycle. 

\begin{figure}[ht!]
    \centering
    \begin{tabular}{|c|c|c|}
    \hline
      $\mu$-chemoEH ($\eps=0.01$) & chemoEH ($\eps=0.01$) & cubic ($\eps=0.01$)  \\
    \hline
    \subfloat[]{\includegraphics[width=.28\textwidth]{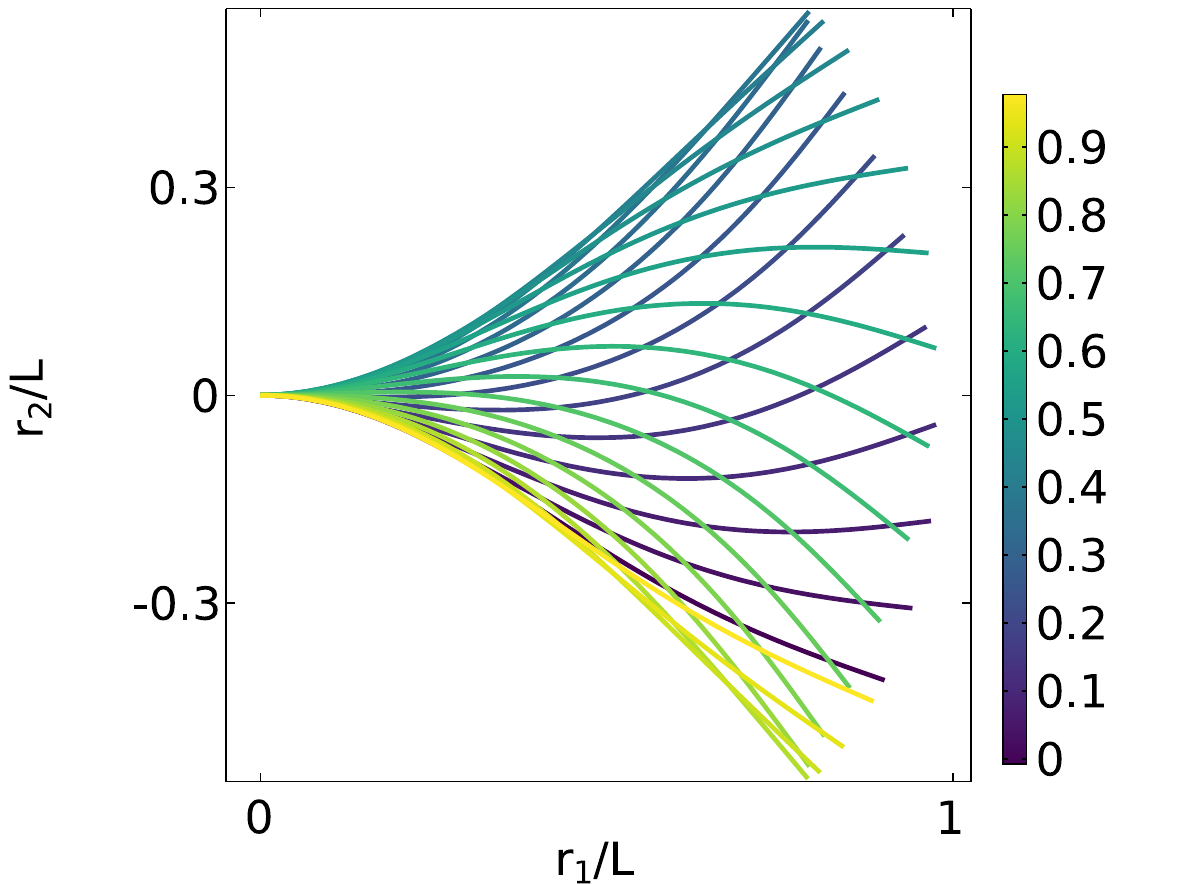}} &
    \subfloat[]{\includegraphics[width=.28\textwidth]{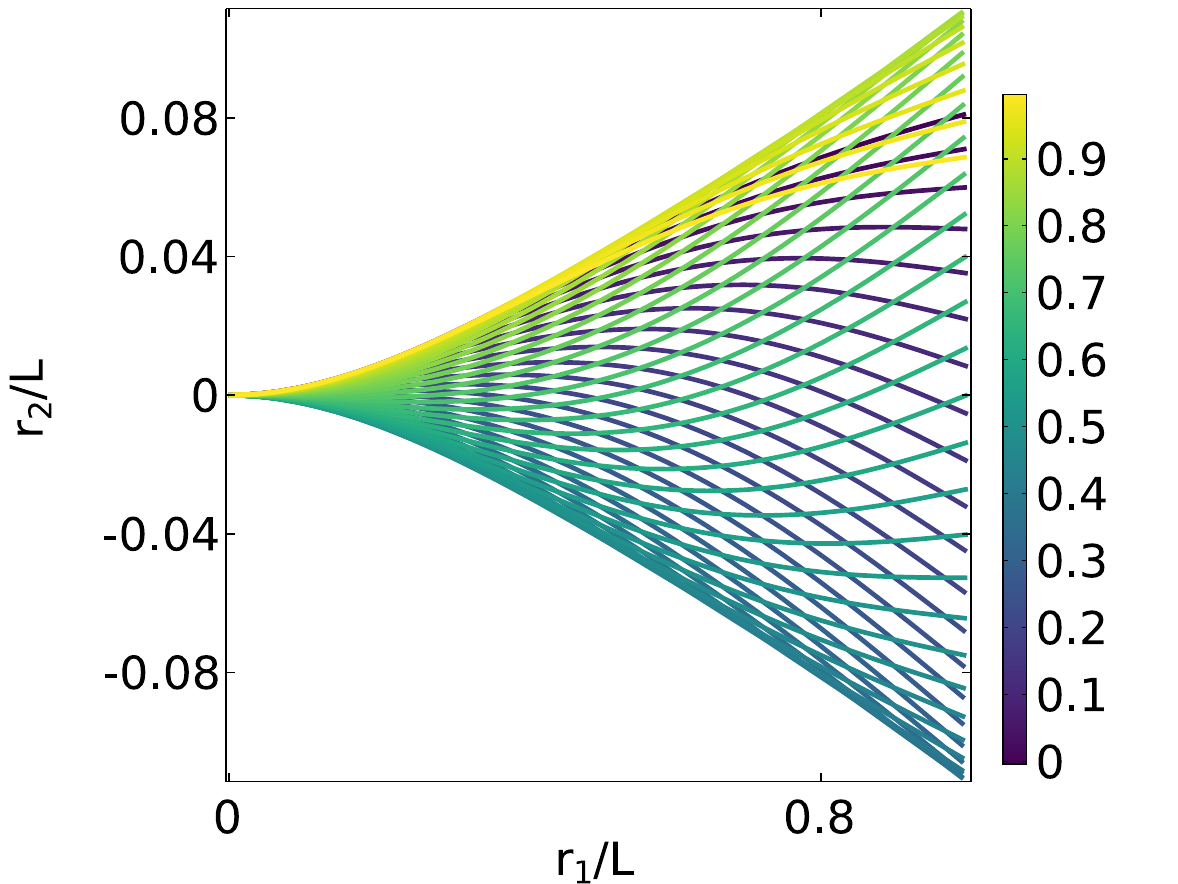}} &
    \subfloat[]{\includegraphics[width=.28\textwidth]{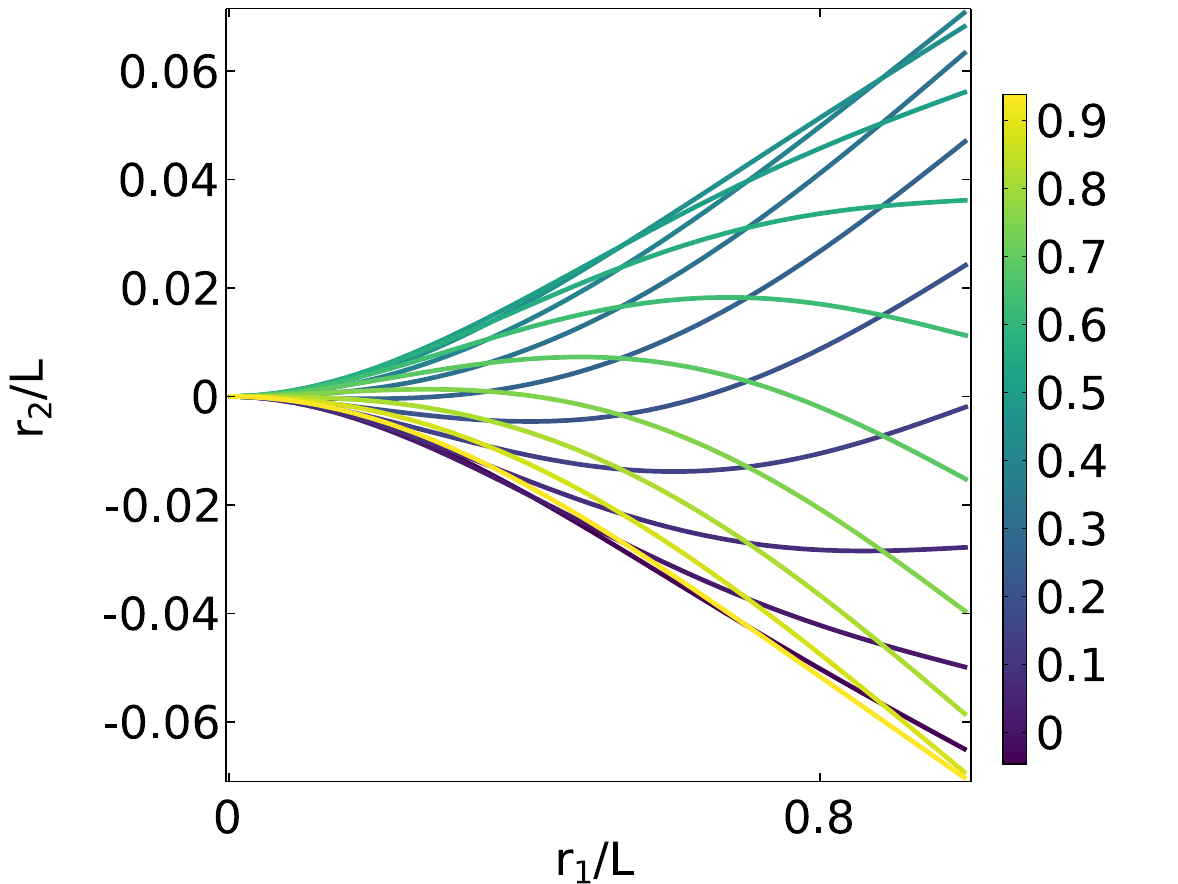}} \\
    \hline
    \subfloat[]{\includegraphics[width=.28\textwidth]{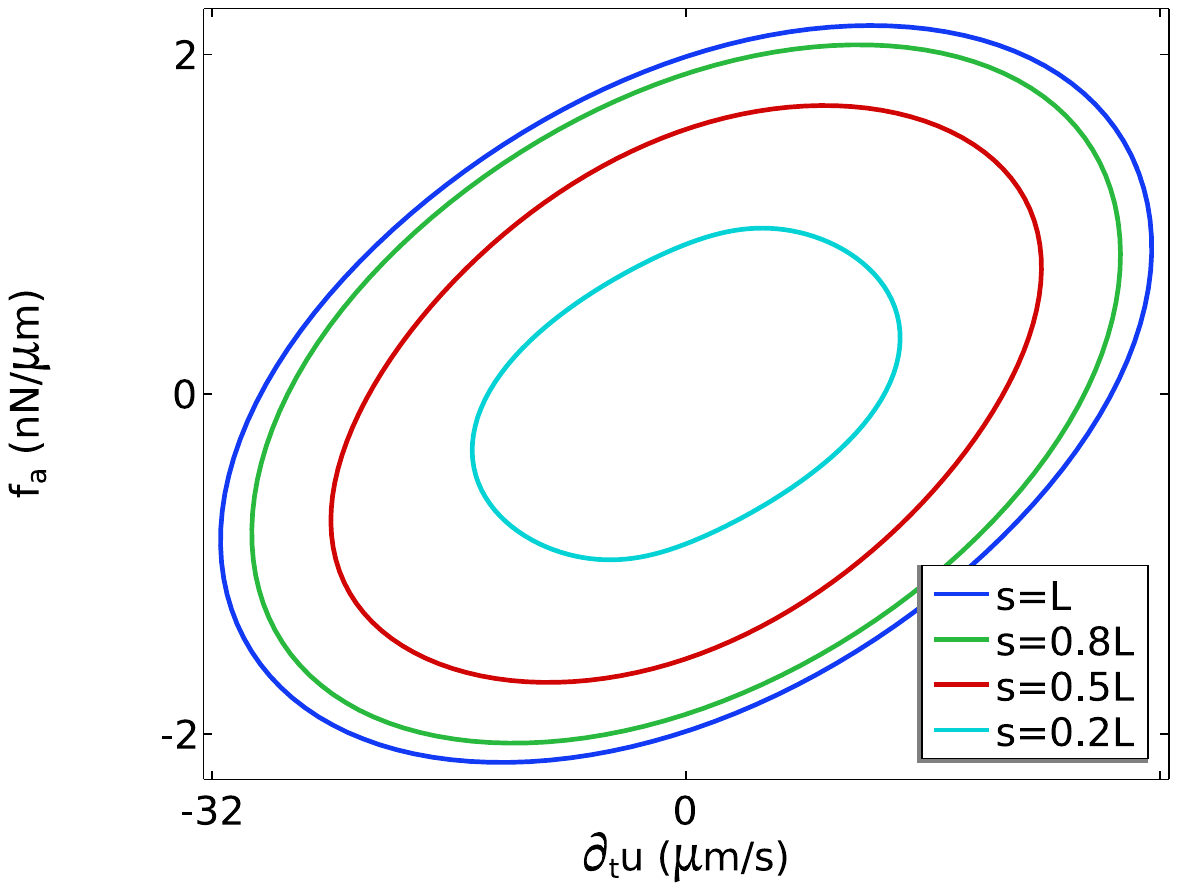}} &
    \subfloat[]{\includegraphics[width=.28\textwidth]{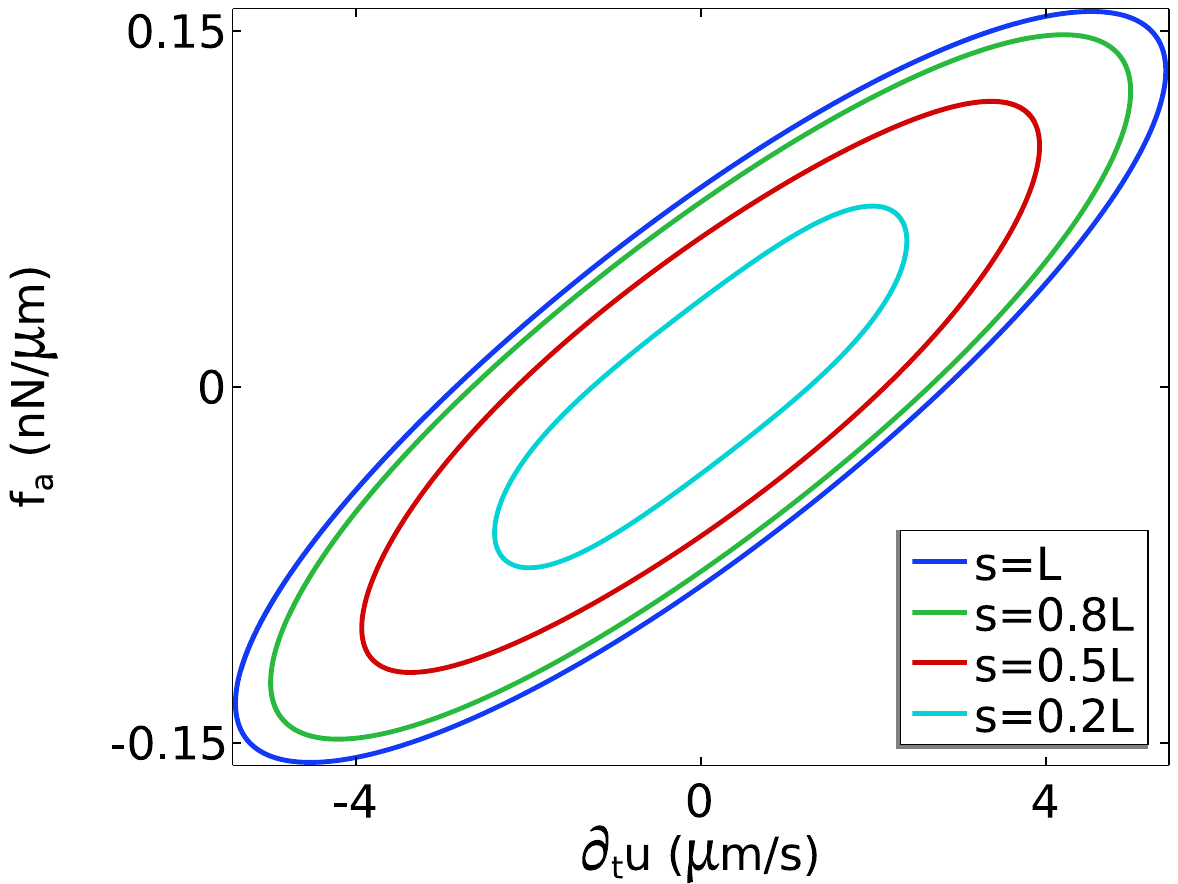}} &
    \subfloat[]{\includegraphics[width=.28\textwidth]{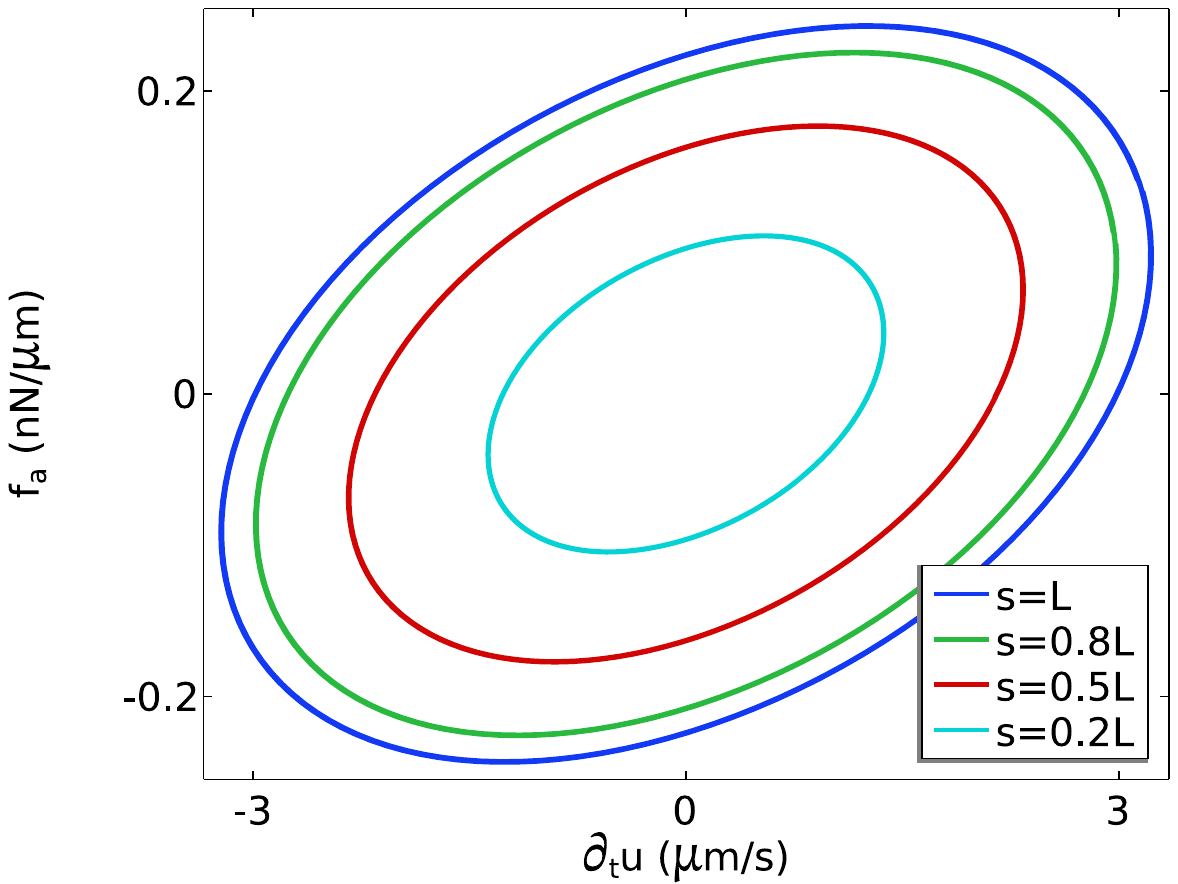}} \\
    \hline
    \subfloat[]{\includegraphics[width=.28\textwidth]{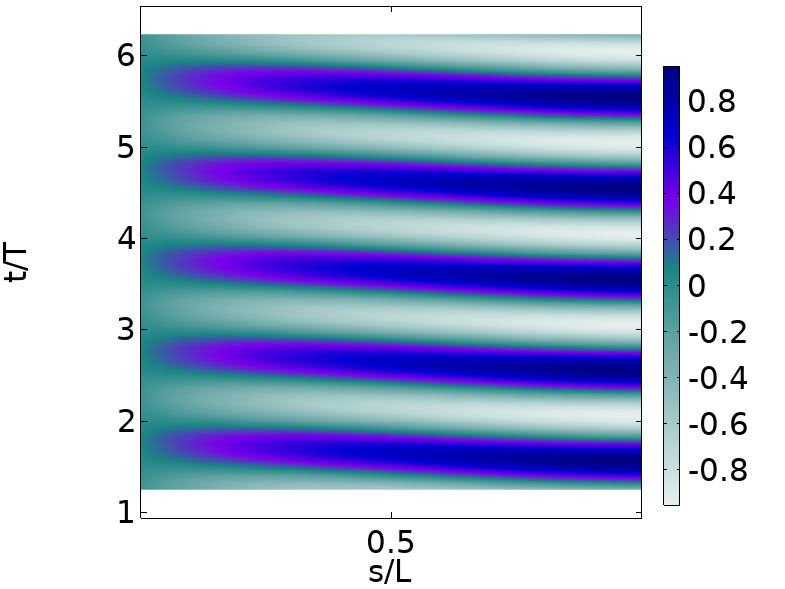}} &
    \subfloat[]{\includegraphics[width=.28\textwidth]{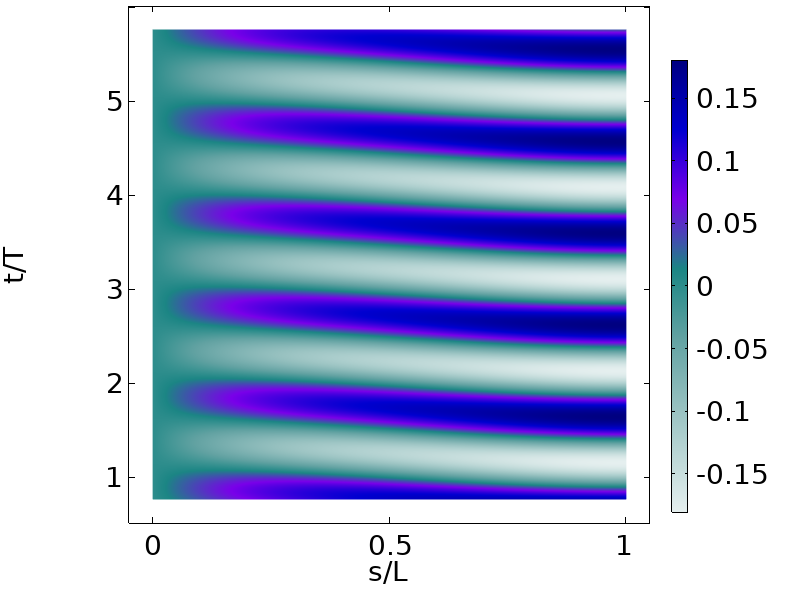}} &
    \subfloat[]{\includegraphics[width=.28\textwidth]{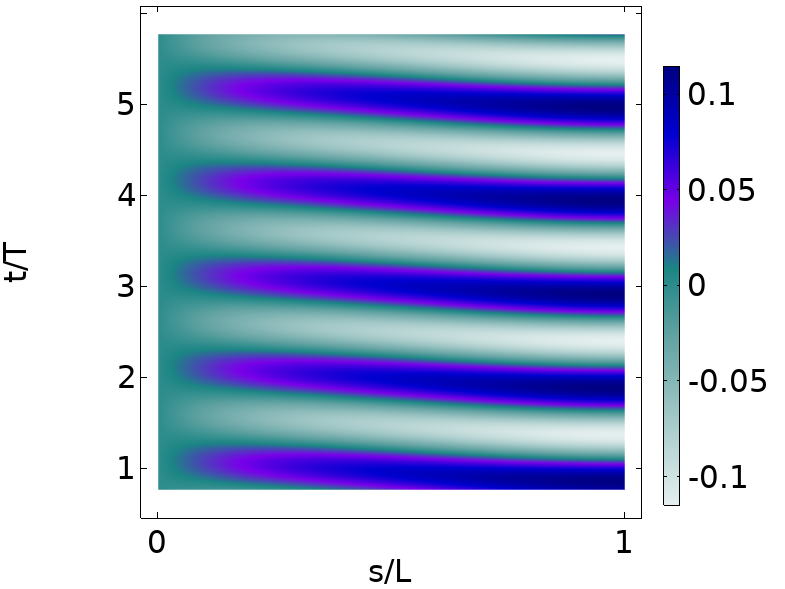}} \\
    \hline
    \end{tabular}
    \caption{Comparison between three models at $\eps=0.01$: $\mu$-chemoEH (first column), chemoEH model (second column), cubic-model (third column). For each model, we show the deformed configuration (first row), the limit cycles between active force and velocity at different arc-length (second row) and the kymograph of the tangent angle $\varphi(s,t)$ with the legend bar on the right showing its amplitude, $s$ is normalized over $L$ and $t$ over a period $T$, (third row).}
    \label{tab:comparison_models_0.01}
\end{figure}

We observe that the $\mu$-chemoEH waveforms, even for small distances $\eps$ from equilibrium, significantly deviate from the straight equilibrium state. Instead, the other two models show less sensitivity to the bifurcation parameter. This is noticeable in Figure~\ref{tab:comparison_models_0.01}(a), where the transverse deformation is almost ten times larger than the ones in Figure~\ref{tab:comparison_models_0.01}(b) and (c). The same observation applies to the maximum amplitude of the active force $ f_a $, where, for both the $\mu$-chemoEH and the cubic model, we considered $ f(t)/2 $ as the total motor force. In the second row of Figure~\ref{tab:comparison_models_0.01}, we plot the active force versus the sliding velocity when the system enters its limit cycle for different arc-lengths, $s \in \{0.2L, 0.5L, 0.8L, L\}$. While they all have  anticlockwise direction, we can observe that the amplitude of the active force, along with the velocity amplitude, is roughly ten times greater for the $\mu$-chemoEH compared to the chemoEH model and the cubic model. The graphs confirm the existence of oscillations around the zero equilibrium point.
The wave forms in the kymograph~\ref{tab:comparison_models_0.01}((g),(h),(i)) are similar, showing wave propagation from tip ($s=L$) to base ($s=0$). Qualitatively, in all three models, the tangent angle $\varphi(s,t)$, deviates little from a sinusoidal shape, with the active force that grows gradually from base to tip. The critical frequencies are the following $\theta_c= 75.6$ Hz for the $\mu$-chemoEH and cubic model, while $\theta_c= 80.2$ Hz for the chemoEH model.

\begin{figure}[ht!]
  \centering
  \begin{tabular}{|c|c|c|}
    \hline
    \textbf{$\mu$-chemoEH ($\eps=0.15$)} & \textbf{chemoEH ($\eps=2$)} & \textbf{Cubic($\eps=1.2$)} \\
    \hline
    \subfloat[]{\includegraphics[width=.28\textwidth]{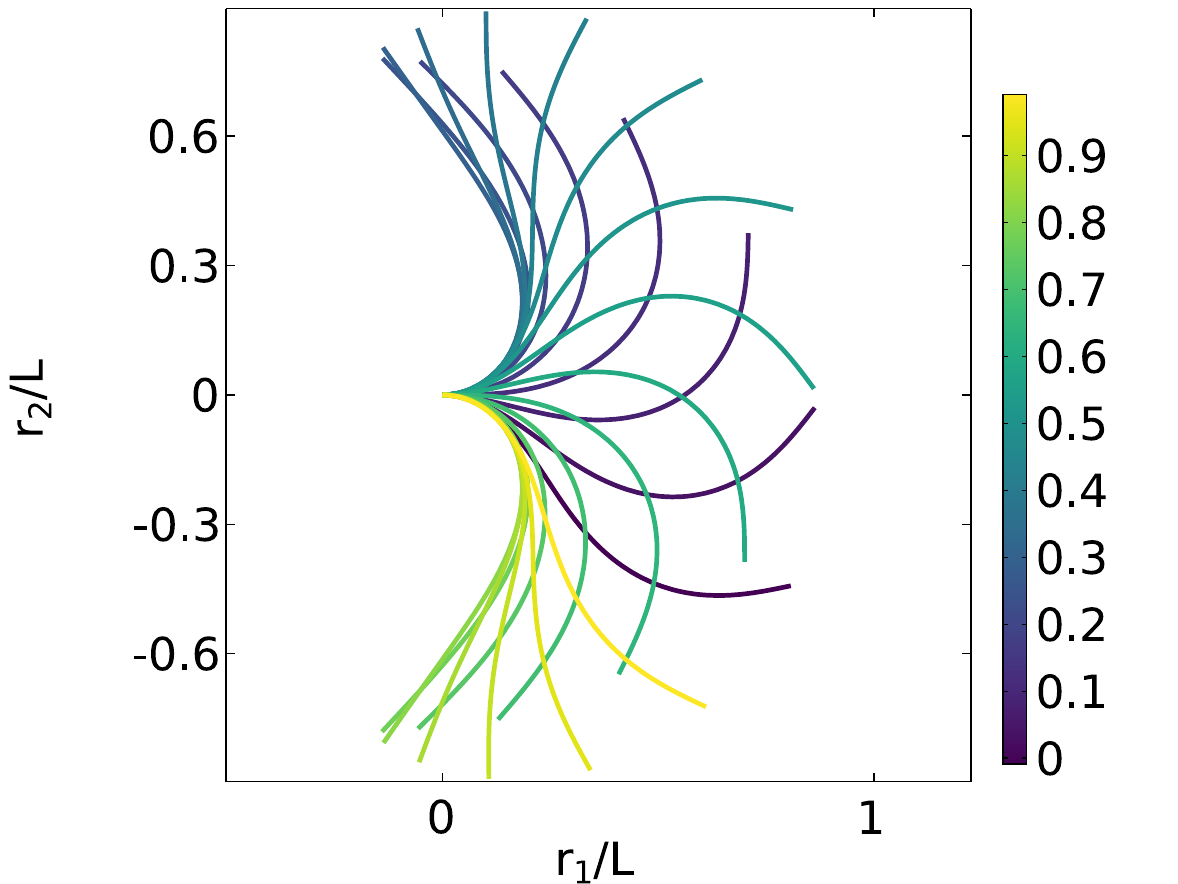}} & \subfloat[]{\includegraphics[width=.28\textwidth]{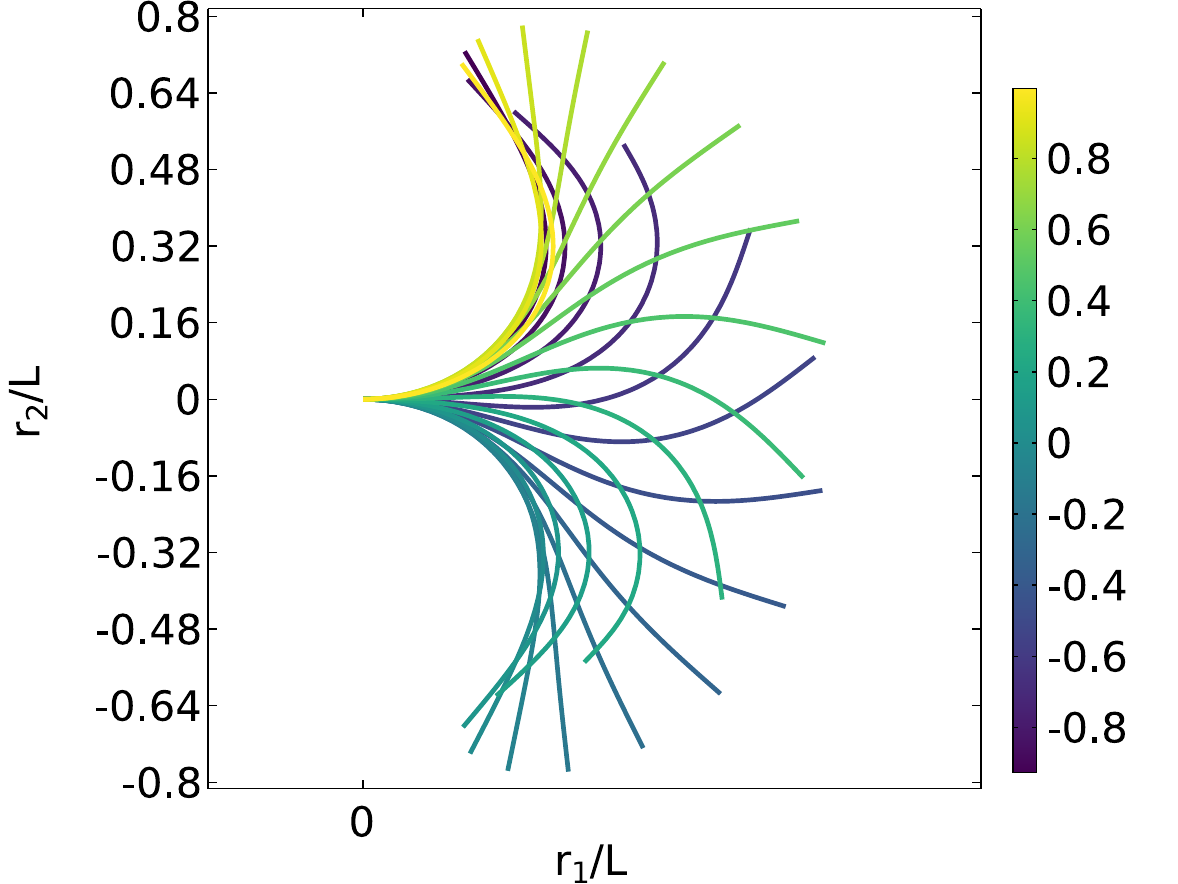}} & \subfloat[]{\includegraphics[width=.28\textwidth]{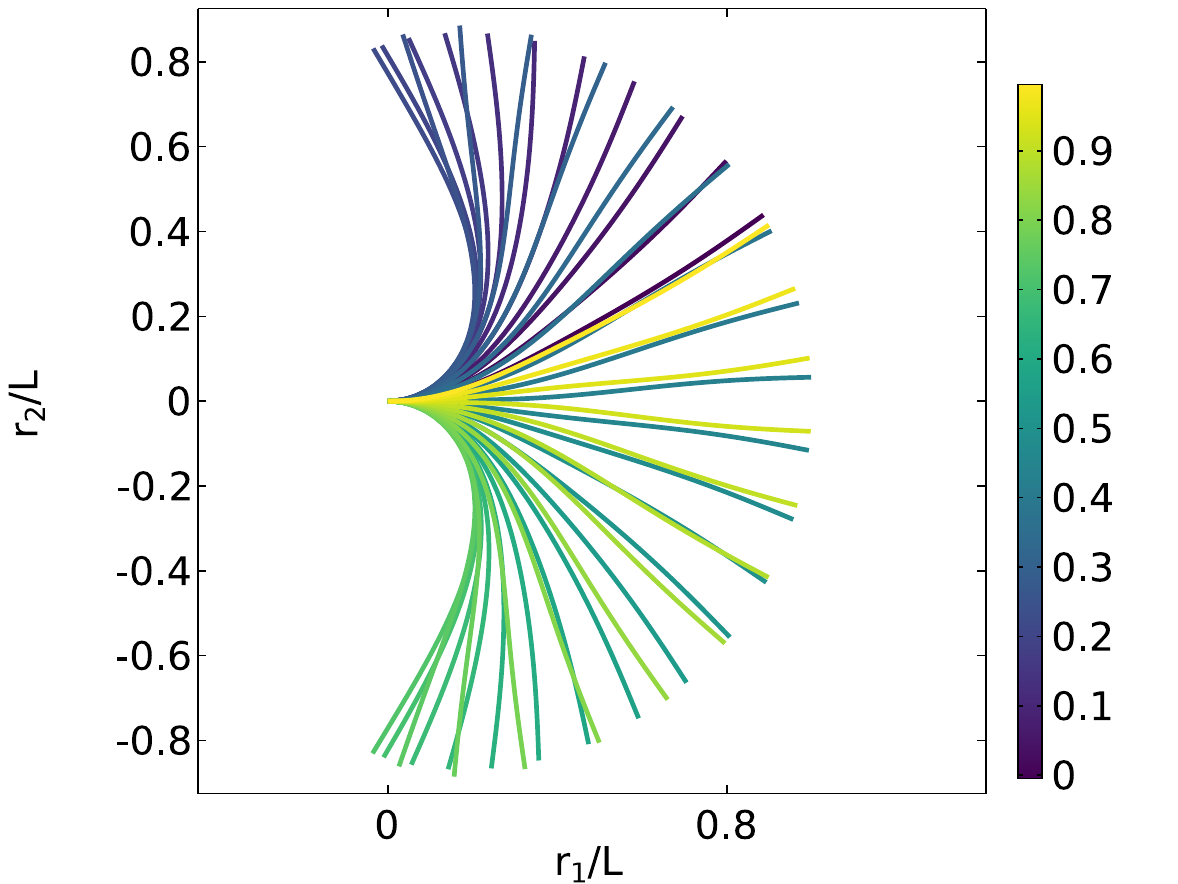}} \\
    \hline
    \subfloat[]{\includegraphics[width=.28\textwidth]{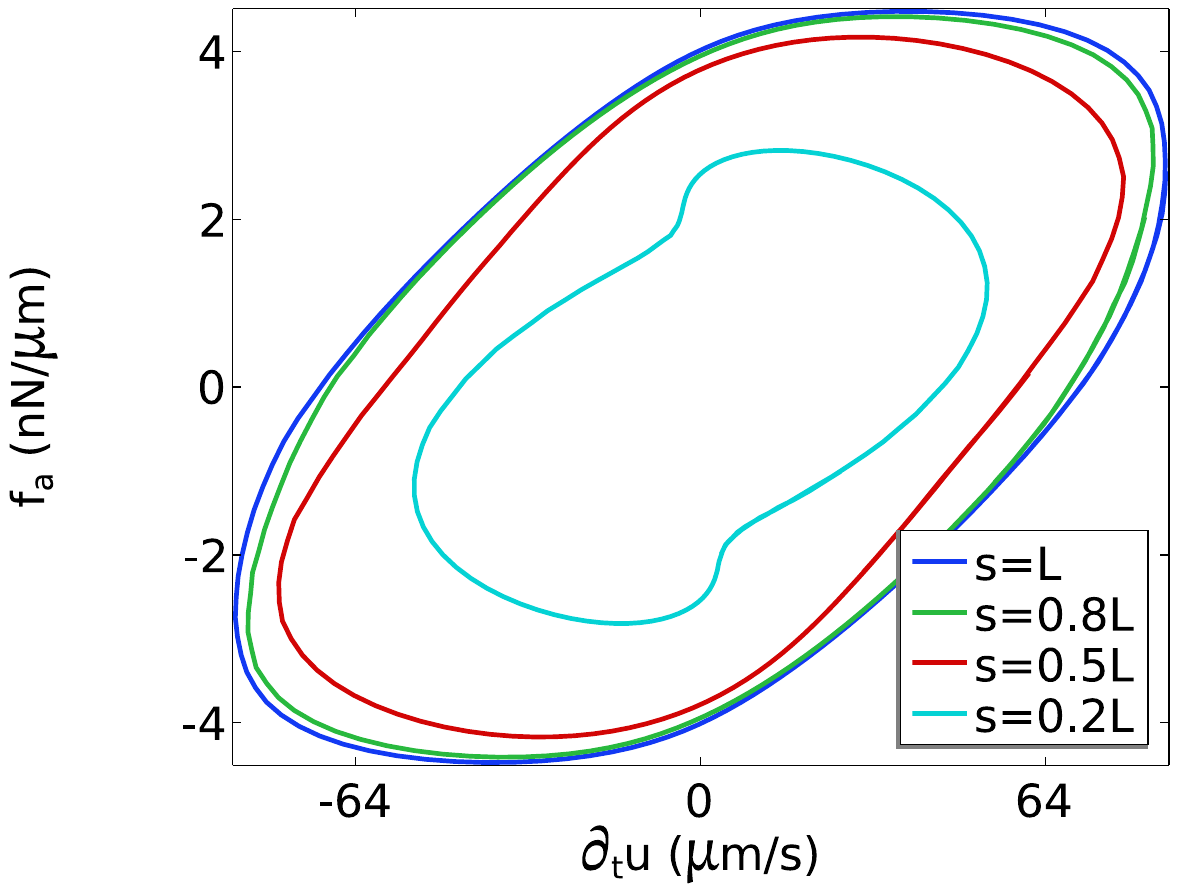}} & \subfloat[]{\includegraphics[width=.28\textwidth]{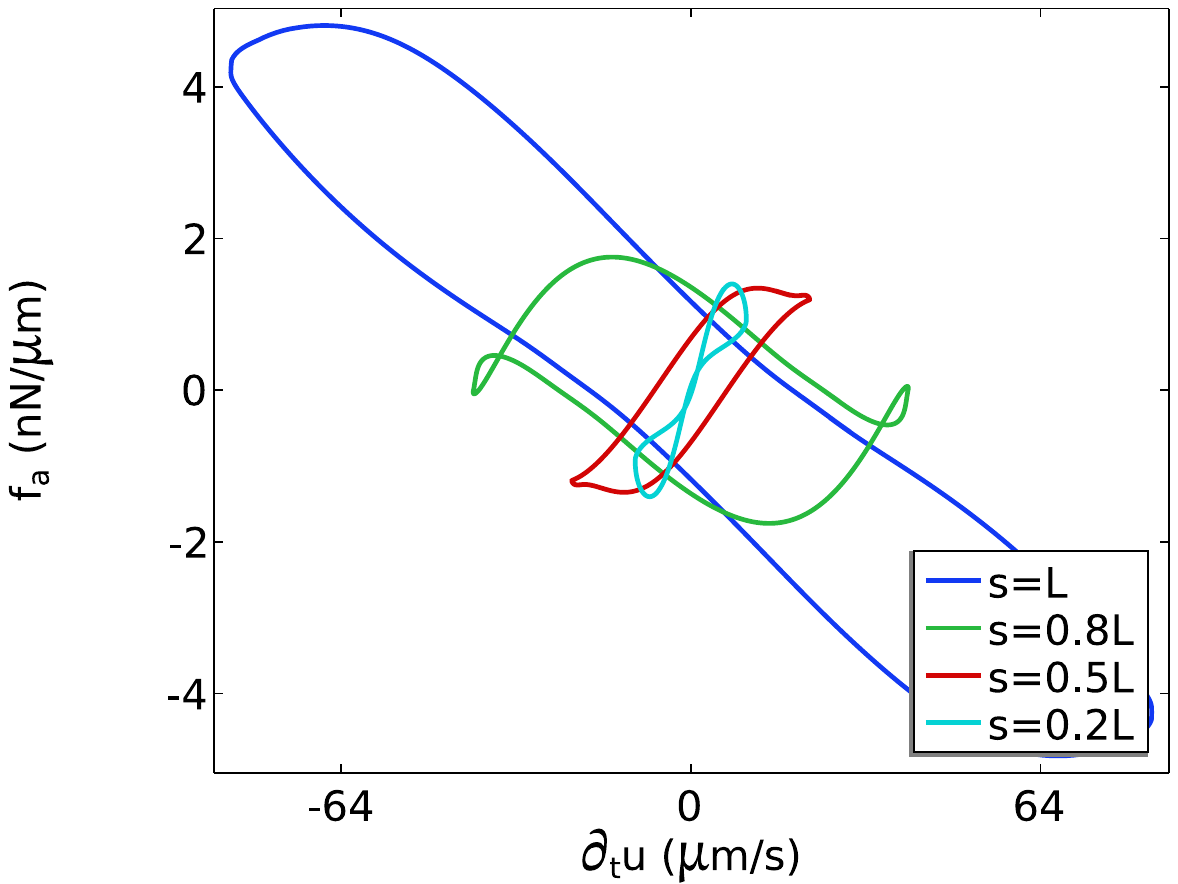}} & \subfloat[]{\includegraphics[width=.28\textwidth]{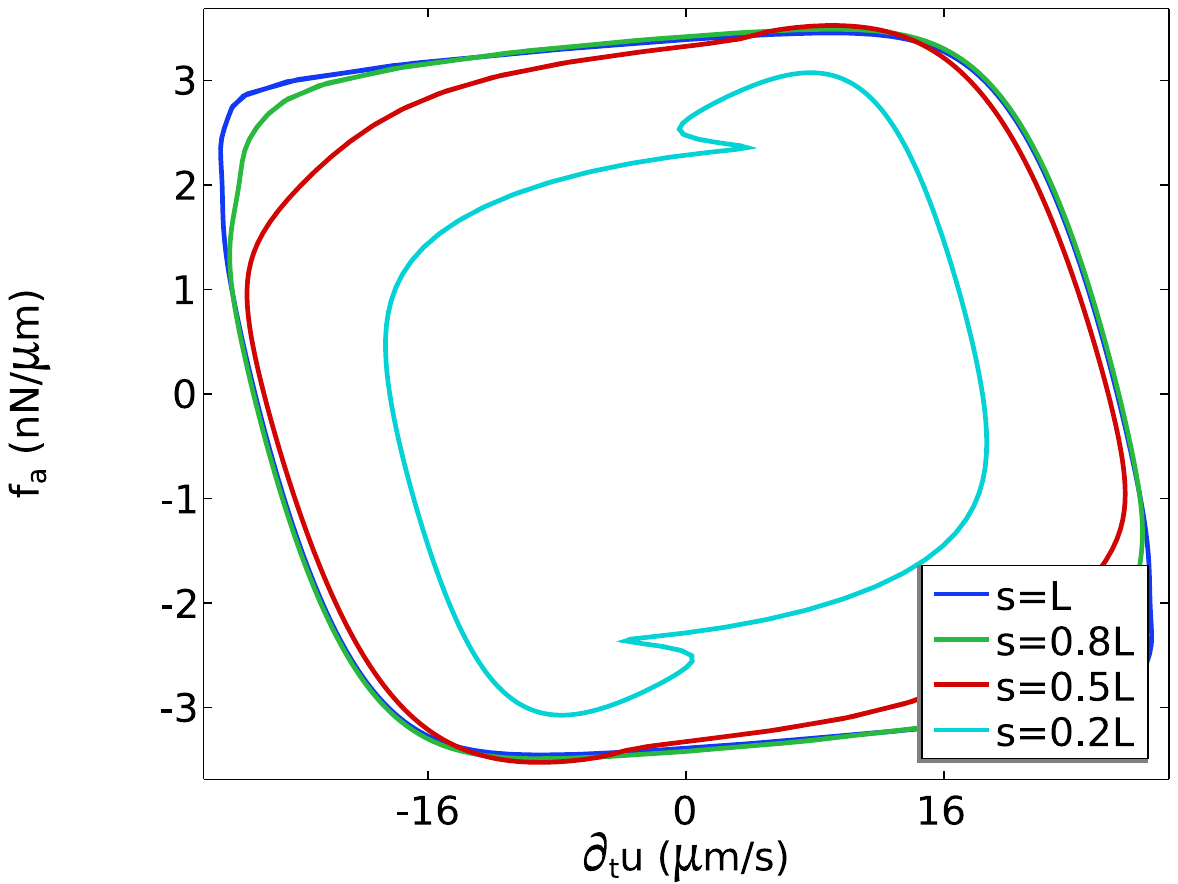}} \\
    \hline
    \subfloat[]{\includegraphics[width=.28\textwidth]{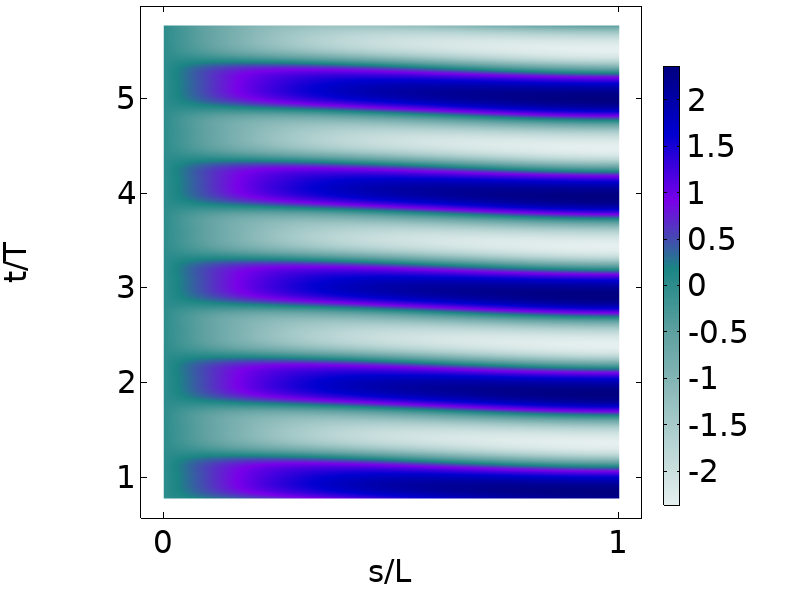}} & \subfloat[]{\includegraphics[width=.28\textwidth]{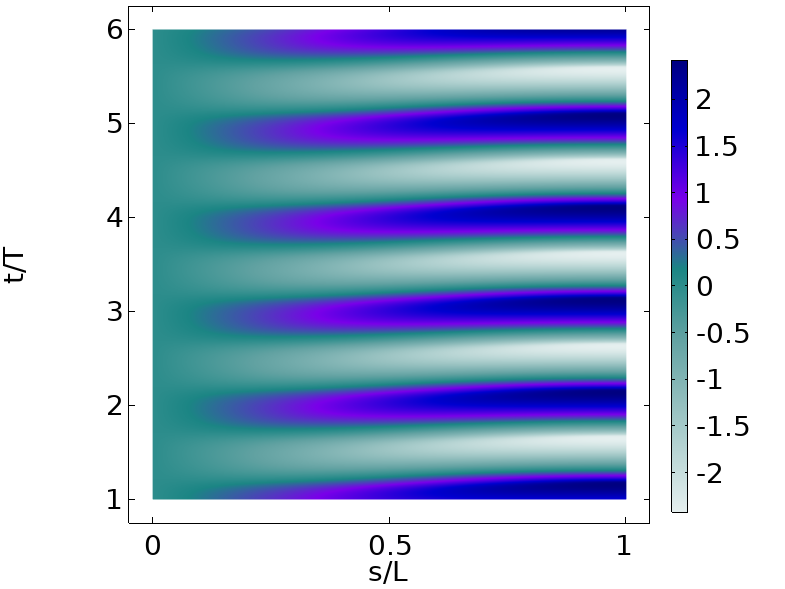}} & \subfloat[]{\includegraphics[width=.28\textwidth]{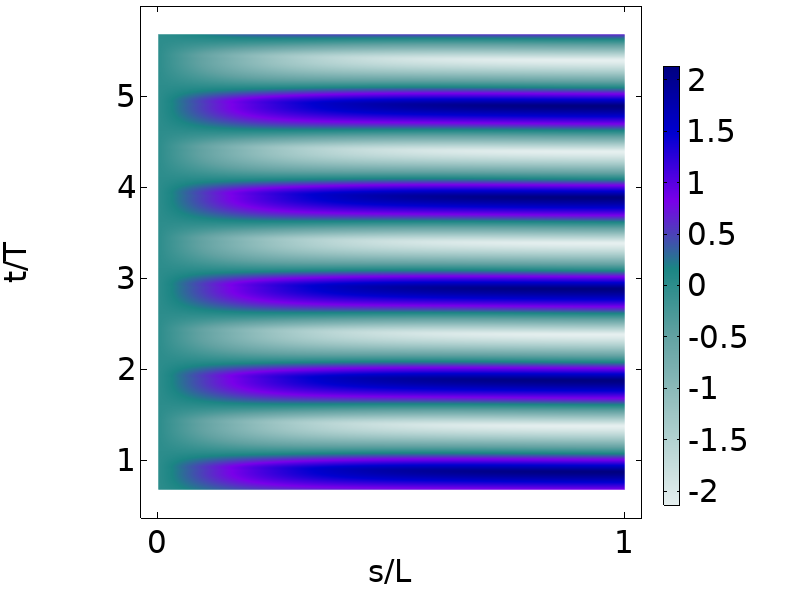}} \\
    \hline
  \end{tabular}
  \caption{Comparison between three models in the non linear regime: same as Figure~\ref{tab:comparison_models_0.01} with $\eps=0.15$ for the $\mu$-chemoEH, $\eps=2$ for the chemoEH model and $\eps=1.2$ for the cubic model.}
  \label{tab:comparison_models_nonlin}
\end{figure}

When increasing the bifurcation value we observe different non linear behaviors depending on the model. Notice that to have comparable deformed configuration, we had to stop at $\eps=0.15$ for the $\mu$-chemoEH, at $\eps=1.2$ for the cubic model and at $\eps=2$ for the chemoEH model, as shown in Figure~\ref{tab:comparison_models_nonlin}. For the $\mu$-chemoEH, even at $\eps=1$, simulations show that the filaments start to twist on themselves.

The deformed configurations plotted in Figures~\ref{tab:comparison_models_nonlin}(a) and Figure~\ref{tab:comparison_models_nonlin}(b) display some similarities in the curvature of the flagellum, while the filaments in Figure~\ref{tab:comparison_models_nonlin}(c) 
are less curved.

We observe that, while in the $\mu$-chemoEH and in the cubic model, the active force versus the sliding velocity curve remains roughly the same starting from half of the flagellum onward, in the chemoEH model the relation looks different depending on the distance from the base. In Figure~\ref{tab:comparison_models_nonlin}(e) we observe that the active force at the tip is twice the one of the rest of the flagella. As in the first group of simulations, near the bifurcation, the direction of the limit cycle is anticlockwise. With this set of parameters, the kymograph in the third row of Figure~\ref{tab:comparison_models_nonlin} show interesting differences. While in the $\mu$-chemoEH model we observe similar velocity and direction of propagation as in the case  $\eps=0.01$, in the chemoEH we observe a change of wave propagation direction, from base to tip, and in the cubic model we recognize the emergence of standing waves. The frequency for the $\mu$-chemoEH stays the same, leading us to notice that some of the linear information were retained by changing $\eps$, probably because the system is still close to the bifurcation parameter. This is not the case for the other two models: in the  chemoEH case, the frequency is $\theta= 40$Hz, while for the cubic model, the frequency decreases to  $\theta= 45.5$Hz. Both the linear and non-linear frequencies lie in the observable range given by \cite{sartori2016dynamic, mondal2020internal}.

\subsection{Tug of war}
Using the parameters in Table~\ref{tab:Chl}, we compute $p^\pm$ and $q^\pm$ as in equation\eqref{eq: P as Fourier series}. These coefficients are then used to obtain the probabilities $P^{\pm}$ within a tug-of-war unit. The interaction between the motors and the filaments, governed by the potentials $\Delta W(\xi)$ generate forces defined as:
\[
f_{\text{pot}}^{\pm}(\xi,t;s,\eps) = P^{\pm}(\xi,t;s,\eps) \partial_{\xi} \Delta W(\xi).
\]
The $\xi$-averaged forces $f_{\text{pot}}^{\pm}$ over the tug-of-war cell are the motor force densities $f^{\pm}(t)$, introduced in equations \eqref{eq:fplus} and \eqref{fmin}. The total active force is then:
\[
f_a = \int_0^\ell \left(f_{\text{pot}}^{+}(\xi) - f_{\text{pot}}^{-}(\xi)\right) \, d\xi.
\]

\begin{figure}[ht!]
  \centering
{\includegraphics[width=1\textwidth]{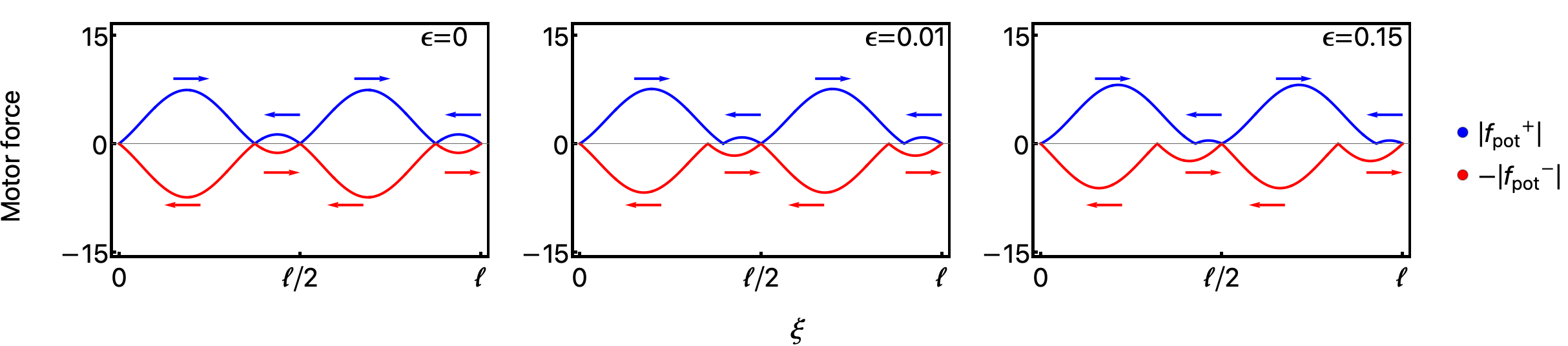}}
  \caption{Tug-of-war between two rows of motors as the bifurcation parameter \(\eps\) varies, with arc length fixed at \(s = 0.8L\). Forces \(|f_{\text{pot}}^{\pm}|\) (blue and red) are shown with arrows indicating their sign. At equilibrium \(\eps = 0\), the system is stalled, with equal and opposite forces. The central picture represents small amplitude oscillations emerging at $\eps=0.01$, while the last picture shows the non linear regime of oscillations, at $\eps=0.15$. Both of these graphs represent snapshots of the oscillations at fixed time $t=400$ ms. In this case both small and large oscillation regimes show greater force on the upper filament than the lower. The non-linear regime also exhibits the highest total active force. Parameters are as in Table~\ref{tab:Chl}.  \label{subplots130}
  }
\end{figure}

Fixing $t=400$ms and $s \in [0,L]$ we examine the system’s behavior when $\eps$ varies. Figure~\ref{subplots130} illustrates this transition, showing the absolute values $|f_{\text{pot}}^{+}|$ and $-|f_{\text{pot}}^{-}|$. Arrows indicate the sign of forces $f_{\text{pot}}^{\pm}$ : positive if pointing to the right and negative otherwise. Without loss of generality, we set $p_0=0$. At the steady state equilibrium, the forces $f_{\text{pot}}^{\pm}$ are equal in magnitude but opposite in direction, implying that motors are stalled with equal stored potential energy. As the chemical energy increases, the forces show different magnitude, leading to a nonzero total active force. This imbalance drives filament sliding, as seen in the plots in Figure~\ref{subplots130} and leads to the so called tug-of-war scenario between motors anchored to opposite filaments. When $\eps = 0.15$, the system enters a nonlinear dynamic and oscillatory regime where one motor group dominates, generating a net active force. 

\begin{figure}[hbt!]
    \centering
\includegraphics[width=0.5\linewidth]{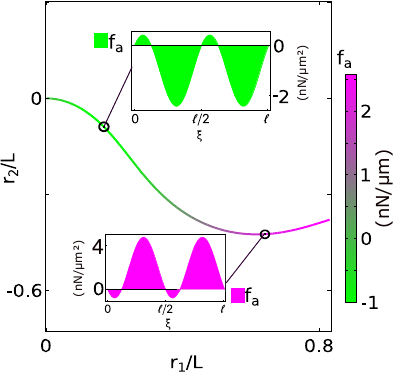}
    \caption{Active force along the deformed configuration at $\eps=0.15$ and time $t=400$ ms. The color bar on the right depicts the active force $f_a$ going from negative, in green, to positive, in magenta.  The graphs of the force $f_{\text{pot}}^{+}-f_{\text{pot}}^{-}$ at $s=0.2L$ and at $s=0.5L$, respectively, are represented in the green and magenta inlets. The area under the graphs is the total active force $f_a$.}
    \label{fig: flagella forces}
\end{figure}

 Figure~\ref{fig: flagella forces} shows the deformed flagellum at $t= 200 $ms,
 with its color representing the active force $f_a(t, s)$ along its length. At $s_1 = 0.2L$ and $s_2 = 0.8L$, the active force have opposite signs. The inlets show the difference $f_{\text{pot}}^{+}-f_{\text{pot}}^{-}$, with areas under the curve representing $f_a(t, s_j)$ for $j=1,2$. Both Figures \ref{subplots130} and \ref{fig: flagella forces} represent the system when the numerical solution has reached its limit cycle.

\newpage

\section{Discussion}
In summary, we have formulated and solved a model coupling the mechanics of a two-dimensional flagellum with clamped-free boundary conditions with the fully nonlinear dynamics of molecular motors. Our model generalizes the two-state model introduced in \cite{camalet2000generic} and we have called this framework the $\mu$-chemoEH model. The issue of having different length scales at play, the macroscopic one given by the filament, and the microscopic one $\xi$, is solved by computing the Fourier expansion for the binding probabilities of the motors. The $\mu$-chemoEH is therefore reduced to a system of five ODEs for the sliding velocity and for only four Fourier coefficients of the probabilities. To obtain this reduction in the number of unknowns, we define both the transition rates and the potentials as cosine and/or sine functions, following  \cite{guerin2011dynamical}. 

In this way, we arrive at a system which is easily comparable with the chemoEH model \cite{oriola2017nonlinear}, from both theoretical and numerical perspectives. The two models present key differences in the hypothesis on the chemistry, in particular on the transition rates, which prevent us to derive the chemoEH model as a simple average of the $\mu$-chemoEH one. We also discuss a third feedback mechanism derived as an approximation of the $\mu$-chemoEH one, termed cubic model, which aligns with the model proposed in \cite{hilfinger2009nonlinear} in the frequency domain. The three models can be summarized as follows: the $\mu$-chemoEH and chemoEH models describe molecular motor dynamics, linking attachment/detachment probabilities, velocity, and active force to oscillations. The $\mu$-chemoEH operates at the microscopic spatial scale $\xi$, while the chemoEH provides averages within a tug-of-war unit. The cubic model focuses solely on the relationship between active force and velocity, embedding microscopic details into the scalar coefficients governing the macroscopic flagellar dynamics.

Simulations confirmed that the $\mu$-chemoEH effectively captures oscillatory behavior near the bifurcation point, which is to be expected since the linear response coefficient \eqref{lin. coeff}
is proportional to the one used in 
\cite{camalet2000generic}. 
The novelty of the $\mu$-chemoEH lies in its nonlinear component, which emerges from the product of the velocity times the Fourier coefficients of the probabilities, as one can appreciate in the ODE system \eqref{eq: Fourier ODE first order p,q,f}. 

The cubic model can be derived from the $\mu$-chemoEH model by approximating the non-linear components of the system in \eqref{eq: Fourier ODE first order p,q,f}. A comparison of the two models, $\mu$-chemoEH and cubic, highlights their differences in handling non-linear components. The 
$\mu$-chemoEH model allows for configurations that are nearly self-coiling under large waveform deformations, whereas the cubic model's non-linearities restrict such behavior. Interestingly, when considering the isolated axoneme $f=0$, the cubic model can be interpreted as a form of van der Pol oscillator for the velocity variable \cite{strogatz2018nonlinear}.

By comparing the three different models, using parameters for short flagella, it is possible to investigate the differences between the nonlinear models far from the bifurcation point by examining, for example, the active force against velocity graphs once the system has entered oscillation (Figure~\ref{tab:comparison_models_nonlin}). Notably, although the magnitudes of force and velocity are comparable across models, the limit cycles reveal that the distribution of active forces along the flagella varies. The two row-model could be an alternative for fitting the data from \textit{Chlamydomonas} and to explore its implications in the case of large amplitude deformations. 

Furthermore, our selected parameters indicate a change in the direction of the emerging traveling waves when moving away from the instability. While the $\mu$-chemoEH model corroborates the existence of the retrograde (tip to base) traveling waves identified by the linear study, the chemoEH model demonstrates a reversal in the sign of propagation, with the waves propagating from base to tip. This propagation is typical of curvature control model, at least in the linear regime \cite{bayly2015analysis}.

We have also obtained feedback graphs (active force vs. velocity or active moment vs. curvature after the system reaches its limit cycle behavior) for the various models analyzed. These graphs are of great importance, since they can serve as a foundation for developing phenomenological models, just as in \cite{gallagher2023axonemal,brokaw1975molecular}.

The $\mu$-chemoEH model makes it possible to study what happens at the microscopic level in a tug-of-war unit by reconstructing the probabilities of detachment and attachment in their entirety. This provides us with a visual representation of the competition between the motors by calculating the active force in the $[0,\ell]$ segment. When the motors stall, non-zero forces oppose each other in sign and direction, creating a zero resultant. When the system receives enough ATP, one motor row prevails over the other, bringing the system out of equilibrium and causing oscillatory sliding between the filaments. 

In conclusion, the $\mu$-chemoEH model fits within the family of nonlinear models based on sliding feedback and aims to describe, in as much detail as possible, the activity of the motors. In the context of a comprehensive elasto-hydrodynamic description of the flagellum, we focus on how the independent molecular motors synchronize and on how the forces generated by the motors coordinate to produce the macroscopic torques that drive the flagellum. It would be interesting to corroborate the finding of this study with experimental data and eventually to reduce the $\mu$ -chemoEH model to a reaction-diffusion model as the ones in \cite{mondal2020internal,cass2023reaction}.  
Also, we have based our $\mu$-chemoEH model on the classical simplified model for the bending elasticity of the axoneme proposed in \cite{camalet2000generic}. Alternative and more detailed/realistic models for the axoneme will be considered in future work, starting from the one proposed in \cite{cicconofri2021biomechanical}.

\section*{Acknowledgments}  We gratefully acknowledge the support by the European Research Council through ERC Advanced Grant MicroMotility (GA 340685) and ERC PoC Grant Stripe-o-Morph (GA 101069436). Additional support was provided by the Italian Ministry of Research through the  projects Response (PRIN 2020) and Abyss (PRIN 2022).

\clearpage
\appendix
\section{On the $\mu$-chemoEH model} 
\subsection{Higher order terms}\label{appendix: tworow-higher order}
By substituting the Fourier expansion \eqref{eq: P as Fourier series} in the PDE system \eqref{eq:PDE motors}, one gets an infinite number of ODEs systems, one for each $n \geq 0$.

For $n = 0$ we obtain two decoupled equations
\begin{equation*}
    \dot{p_0}^\pm(t) = - \alpha p_0^\pm(t) + \alpha \eta / \ell,
\end{equation*}
whose solution exponentially decay to $\eta/\ell$ for large times.
For $n=1$ we get \eqref{eq: Fourier ODE first order p,q,f}, while for $n>1$, we obtain
\begin{equation} \label{eq:appendix pq}
    \left\{\begin{array}{l}
\dot{p}_n^+(t) + \frac{2 \pi}{\ell} \dot{u}(t) q_n^+(t)= - \alpha p_n^+(t),\\\\
\dot{q}_n^+(t) - \frac{2 \pi}{\ell} \dot{u}(t) p_n^+(t)= - \alpha q_n^+(t), \\\\
\dot{p}_n^-(t) - \frac{2 \pi}{\ell} \dot{u}(t) q_n^-(t)= - \alpha p_n^-(t) ,\\\\
\dot{q}_n^-(t) + \frac{2 \pi}{\ell} \dot{u}(t) p_n^-(t)= - \alpha q_n^-(t)
\end{array}\right.
\end{equation}
The solutions $q_n^\pm$ and $p_n^\pm$ with $n>1$ are completely determined by knowing $u(t)$, and they go to zero for large times, independently on the initial conditions. We therefore omit to explicitly solve this last system, and just consider $p_n^\pm=q_n^\pm=0$ when $n>1$.

\subsection{Isolated axoneme}\label{appendix: tworow-higher isolated axoneme}
Now we study the isolated motor-filament systems for the ODE \eqref{eq: Fourier ODE first order p,q,f} by letting $f(t)=0$. We focus on the linearized problem, while a comprehensive non linear analysis is carried out in  \cite{alouges2024mathematicalmodelsflagellaractivation}. We rewrite the system as
\begin{equation} \label{eq_isolated: Fourier ODE first order p,q,f=0}
    \left\{\begin{array}{l}
\dot{p}^+(t) =- \frac{2 \pi}{\ell} \dot{u}(t) q^+(t)- \alpha p^+(t) + \alpha \Omega / (2 \pi^2 \ell),\\\\
\dot{q}^+(t) =\frac{2 \pi}{\ell} \dot{u}(t) p^+(t) - \alpha q^+(t) + \alpha \Omega / (2 \pi^2 \ell),\\\\
\dot{p}^-(t) =\frac{2 \pi}{\ell} \dot{u}(t) q^-(t)- \alpha p^-(t) + \alpha \Omega / (2 \pi^2 \ell),\\\\
\dot{q}^-(t) = - \frac{2 \pi}{\ell} \dot{u}(t) p^-(t)- \alpha q^-(t) + \alpha \Omega / (2 \pi^2 \ell),\\\\
\dot{u}(t) = \frac{1}{2 \lambda}\left(- 2 K u(t) +  \rho N \pi U \left(q^+(t) - q^-(t)\right)\right).
\end{array}\right.
\end{equation}

The Jacobian of this system, computed at the equilibrium point $p^\pm_{eq}=q_{eq}^\pm= \Omega/ (2 \pi^2 \ell)$ and $u_{eq}=0$, reveals five eigenvalues $\lambda_i$: the first three are negative, $\lambda_i=-\alpha$ for $i= {1,2,3}$, while the other two are complex and conjugated $\lambda_{4,5} = \tau \pm i \omega$, where $ \tau(\Omega)= -\frac{K}{2\lambda}-\frac{\alpha}{2}+ \Omega \frac{U \rho N}{2 \ell^2 \lambda}$ and $\omega(\Omega)= \sqrt{\frac{K \alpha}{\lambda}-\tau^2(\Omega)}$.

Furthermore, there exist a critical ATP concentration value $\Omega_0= \ell^2(K\alpha + \lambda)/ ( U \rho N)$ for which $\tau(\Omega_0)=0$ and $\omega(\Omega_0) =\sqrt{\frac{K \alpha}{\lambda}} 
 \neq 0$. In particular, by varying $\Omega$, $\tau=\tau(\Omega)$ goes from negative to positive when crossing $\Omega_0$. We are then in the right hypothesis to reduce the system \eqref{eq_isolated: Fourier ODE first order p,q,f=0} to a two-dimensional system that shows an Hopf-bifurcation around the equilibrium point and near $\Omega_0$. We transform the system in such a way that the linear matrix associated to \eqref{eq_isolated: Fourier ODE first order p,q,f=0} becomes a block matrix. The transformation is the following: at first we bring the system to the origin
\begin{equation*}
    \delta p^\pm = p^\pm - p_{eq}^\pm,\,\delta q^\pm = q^\pm - q_{eq}^\pm,
\end{equation*}
then we write
\begin{equation*}\label{eq: change of variable th}
    \begin{array}{cccc}
         r = \delta p^{+} + \delta q^+ & 
         s = \delta p^- +\delta q^-, &
         z = \frac{1}{2}(\delta q^+ + \delta q^-), & 
         y = \frac{1}{2}(\delta q^+ - \delta q^-).
    \end{array}
\end{equation*} 
In this way, the linear part of the system \eqref{eq_isolated: Fourier ODE first order p,q,f=0} becomes
\begin{equation*} 
\left(\begin{array}{l}
\dot r\\
\dot s\\
\dot z\\
\end{array}\right) = -\alpha\left(\begin{array}{l}
r\\
s\\
z\\
\end{array}\right),\quad \left(\begin{array}{l}
\dot y\\
\dot u
\end{array}\right) = \frac{1}{\lambda}\left(
\begin{array}{cc}
  -\alpha \lambda + k_{cb}\Omega \rho N & - \frac{\Omega K}{\pi \ell^2}  \\
 \rho N \pi U & -K\\
\end{array}
\right)\left(\begin{array}{l}
y\\
u
\end{array}\right)
\end{equation*}
At the critical bifurcation point $\Omega_0$ the variables $r$, $s$ and $z$ decay to their equilibrium point in time, while $y$ and $u$ oscillates around zero with frequency $\omega_0=\omega(\Omega_0)$. This means that the pair $(y,u)$ is the one that contributes to the Hopf-bifurcation. In order to recover the information for the original variables, we transform back the system

and get, at a linear level, that all the variables $p^\pm$, $q^\pm$ and $u$ oscillate around their own equilibrium position with frequency $\omega_0$. 

In conclusion, while the ATP concentration is such that $\Omega \ll \Omega_0$, the motors are stalled at the equilibrium, which depends on $\Omega$, and the filament does not move. In this regime, molecular motors are storing chemical energy, which is then converted it into mechanical work as soon as the ATP concentration becomes greater than the critical value $\Omega_0$. 

\section{On the chemoEH model} \label{appendix: chemoEH}
We can compare the $\mu$-chemoEH with the chemoEH by considering the last with zero force $f(t)=0$
\begin{equation}\label{eq: ode n+n-f=0}
    \left\{\begin{array}{ll}
 \dot{n}_+= -\pi_0 (N-n^+) -\frac{\eps_0}{2}e^{\bar f} n^+ e^{- \bar f \dot u / v_0} \\
    \dot{n}_-= -\pi_0 (N-n^-) -\frac{\eps_0}{2}e^{\bar f} n^- e^{ \bar f \dot u / v_0}\\
     0=  (n^+-n^-) - (n^+ + n^-)  \dot u / v_0-\frac{K}{f_0 \rho} u.
    \end{array}\right.
\end{equation}
First, we observe that the equilibria are $n^{\pm}_{eq} = n_0 = \pi_0 N / \alpha_0$ and $u^{eq} = 0$. We then apply two invertible transformations: first, defining $\delta n^{\pm} = n^{\pm} - n_0$, and subsequently, defining $z = (\delta n^+ + \delta n^-)/2$ and $y = (\delta n^+ - \delta n^-)/2$. After these transformations, the linearized system becomes
\begin{equation} \label{eq: 5d compact th general}
\dot{z} = -\alpha_0 z, \quad \left( \begin{array}{l} 
\dot{y} \\ 
\dot{u} 
\end{array} \right) = \left( 
\begin{array}{cc} 
-\alpha_0 - \eps_0 e^{\bar{f}} \bar{f} & \eps_0 e^{\bar{f}} \bar{f} \frac{K}{2 f_0 \rho} \\
\frac{v_0}{n_0} & -\frac{K v_0}{2 n_0 f_0 \rho} 
\end{array} 
\right) \left( \begin{array}{l} 
y \\ 
u 
\end{array} \right).
\end{equation}

When restricting the dynamics to the plane $(y, u)$, we observe the onset of a Hopf bifurcation, as shown in~\ref{appendix: tworow-higher isolated axoneme} for the $\mu$-chemoEH system. In fact, the system has three eigenvalues: the first, $-\alpha_0$, is always negative, while the other two, $\lambda_{2,3} = \tau \pm i \omega$, are complex conjugates, where
\[
\tau = -\left( \frac{K v_0}{2 n_0 f_0 \rho} + \frac{\alpha_0}{2} - \frac{1}{2} \eps_0 e^{\bar{f}} \bar{f} \right)
\]
and
\[
\omega = \sqrt{K \alpha_0 \frac{v_0}{2 n_0 f_0 \rho} - \tau^2}.
\]
Moreover, we observe that $\tau(f_0) = 0$ when 
\[
f_0 = v_0 \frac{K(\pi_0 + \eps_0 e^{\bar{f}})}{2 N \pi_0 \rho (-\pi_0 + \eps_0 e^{\bar{f}}(\bar{f} - 1))}.
\]
In general, $\tau(f_0)$ changes sign as $f_0$ varies, making $f_0$ the bifurcation parameter of the system in this case. At the linear level, the two models are equivalent.  

\section{Numerical simulations methods} \label{app:numerics}
The governing equations are implemented with the finite element software COMSOL Multiphysics in the equation mode, using a time-dependent environment. For each model, the torque balance equation \eqref{eq: local moment balance} is coupled with system \eqref{eq: microscopic system}, implemented via the weak form PDE and the ODE interface, respectively. The weak form of the torque balance equation 
reads
\begin{equation*}
    -B\int_0^L \varphi'(s,t) \alpha'(s,t) \,ds + \int_0^L (T(s,t) + a f(s,t))\alpha(s,t) \,ds =0\,,
\end{equation*}
where $\alpha$ is a test function with the same clamped-free boundary conditions as $\varphi$.
The BDF solver is used for the time-stepping. A Newton algorithm is employed to iteratively solve the non-linear algebraic system resulting
from the finite element discretization at each time step. The direct
solver MUMPS is chosen for the solution of the linearized system at
each iteration.

\printbibliography
\end{document}